\documentstyle[12pt,aasms4]{article}

\lefthead{Gonzalez et al.}
\righthead{Abundance Analyses of 20 New Systems}
\slugcomment{Submitted to the Astronomical Journal}
\received{}
\accepted{}
\begin{document}

\title{Parent Stars of Extrasolar Planets VI: Abundance Analyses of 20 New 
Systems}

\author{Guillermo Gonzalez\altaffilmark{1}, 
Chris Laws\altaffilmark{1}, Sudhi Tyagi\altaffilmark{1}, and 
B. E. Reddy\altaffilmark{2}}

\altaffiltext{1}{University of Washington, Astronomy Department, Box 351580, 
Seattle, WA 98195 gonzalez@astro.washington.edu, laws@astro.washington.edu, 
styagi@u.washington.edu}
\altaffiltext{2}{Department of Astronomy, University of Texas, Austin, TX 
78713-1083 ereddy@shaka.as.utexas.edu}

\begin{abstract}

The results of new spectroscopic analyses of 20 recently reported extrasolar 
planet parent stars are presented.  The companion of one of these stars, 
HD\,10697, has recently been shown to have a mass in the brown dwarf regime; 
we find [Fe/H] $= +0.16$ for it.  For the remaining sample, we derive [Fe/H] 
estimates ranging from $-0.41$ to $+0.37$, with an average value of $+0.18 
\pm 0.19$.  If we add the 13 stars included in the previous papers of this 
series and 6 other stars with companions below the 11 M$_{\rm Jup}$ limit 
from the recent studies of Santos et al., we derive $\langle$[Fe/H]$\rangle = 
+0.17 \pm 0.20$.

Among the youngest stars with planets with F or G0 spectral types, [Fe/H] 
is systematically larger than young field stars of the same Galactocentric 
distance by 0.15 to 0.20 dex.  This confirms the recent finding of Laughlin 
that the most massive stars with planets are systematically more metal rich 
than field stars of the same mass.  We interpret these trends as supporting a 
scenario in which these stars accreted high-Z material after their convective 
envelopes shrunk to near their present masses.  Correcting these young star 
metallicities by 0.15 dex still does not fully account for the difference in 
mean metallicity between the field stars and the full parent stars sample.

The stars with planets appear to have smaller [Na/Fe], [Mg/Fe], and [Al/Fe] 
values than field dwarfs of the same [Fe/H].  They do not appear to have 
significantly different values of [O/Fe], [Si/Fe], [Ca/Fe], or [Ti/Fe], 
though.  The claim made in Paper V that stars with planets have low [C/Fe/] 
is found to be spurious, due to unrecognized systematic differences among 
published studies.  When corrected for these differences, they instead display 
slightly enhanced [C/Fe] (but not significantly so).  If these abundance 
anomalies are due to the accretion of high-Z matter, it must have a 
composition different from that of the Earth.

\end{abstract}


\section{INTRODUCTION}
\label{intro}

In our continuing series on stars-with-planets (hereafter, SWPs), we have 
reported on the results of our spectroscopic analyses of these stars 
(Gonzalez 1997, Paper I; Gonzalez 1998, Paper II; Gonzalez \& Vanture 1998, 
Paper III; and Gonzalez et al. 1999, Paper IV; Gonzalez \& Laws 2000, 
Paper V).  Other similar studies include Fuhrmann et al. (1997, 1998) and 
Santos et al. (2000b,c).  The most significant finding so far has been the 
high mean metallicity of SWPs, as a group, compared to the metallicity 
distribution of nearby solar-type stars (Gonzalez 2000; Santos et al. 
2000b,c).

Additional extrasolar planet candidates continue to be announced by planet 
hunting groups using the Doppler method.  We follow-up these annoucements 
with high resolution spectroscopic observations as time and resources 
permit.  Herein, we report on the results of our abundance analyses of 20 
new candidate SWPs.  We compare our findings with those of other recent 
similar studies, look for trends in the data suggested in previous studies, 
and evaluate proposed mechanisms in light of the new dataset.

\section{SAMPLE AND OBSERVATIONS}
\label{obs}

High-resolution, high S/N ratio spectra of 14 stars were obtained with the 
2dcoude echelle spectrograph at the McDonald observatory 2.7 m telescope 
using the same setup as described in Paper V.  Two stars difficult or 
impossible to observe from the northern hemisphere, HR 810 and HD 1237, were 
observed on three nights with the CTIO 1.5 m with the fiber fed echelle 
spectrograph.  Observing them on multiple nights permits us to test for 
possible variations in their temperatures over one stellar rotation period, 
given their youth.  Additional details of the spectra obtained at CTIO and 
McDonald, including a list of the discovery papers, are presented in Table 
1.  Although it does not have a known planet, we include HD\,75332 in the 
program, since its physical parameters are similar to those of the hotter 
SWPs.  HD\,75332 is also included in the field star abundance survey of Chen 
et al. (2000), which we will be comparing to our results in Section 4.2.7.  We 
also include HD\,217014 (51 Peg), even though it was already analyzed in Paper 
II, because: 1) the new spectra are of much higher quality, and 2) it was 
included in the field star abundance surveys of Edvardsson et al. (1993) and 
Tomkin et al. (1997). 

High resolution spectra of nine stars (HD\,12661, HD\,16141, HD\,37124, 
HD\,38529, HD\,46375, HD\,52265, HD\,92788, HD\,177830, and BD\,-10 3166)
\footnote{The discovery papers corresponding to these stars are: Butler et al. 
2000; Fischer et al. 2000; Marcy et al. 2000; Sivan et al. 2000; Vogt et al. 
2000.} obtained with the HIRES spectrograph on the Keck I were supplied to us 
by Geoff Marcy (see Paper IV for more details on the instrument).  The Keck 
spectra have the advantage of higher resolving power and much weaker water 
vapor telluric lines, due to the altitude of the site.  However, the much 
smaller wavelength coverage of the Keck spectra results in a much shorter 
linelist for us to work with.

The data reduction methods are the same as those employed in Paper V.  Spectra 
of hot stars with a high $v \sin i$ values were also obtained in order to 
divide out telluric lines in the McDonald and CTIO spectra.

\section{ANALYSIS}
\subsection{Spectroscopic Analysis}

The present method of analysis is the same as that employed in Paper V, and 
therefore, will not be described herein.  We have added more Fe I, II 
lines to our linelist (Table 2).  Their $gf$-values were calculated from 
an inverted solar analysis using the Kurucz et al. (1984) Solar Flux Atlas 
or our spectrum of Vesta (obtained with the McDonald 2.7 m).  We also added 
a new synthesized region: 9250 - 9270 \AA.  This region contains one Mg I, 
two Fe I, and three O I lines; only one of the O I triplet, 9266 \AA, is 
unblended in all our stars, but the other two are usable in the warmer 
stars.  The addition of a second Mg line to our linelist helps greatly, 
because the 5711 line was the only one we had employed until now, and it is 
not measurable in the cooler stars.  We also added the O I triplet near 7770 
\AA.  Since these lines are known to suffer from non-LTE effects, we have 
corrected the O abundances derived from these lines using Takeda's (1994) 
calculations.  We list the individual EW values in Tables 3 - 6 and present 
the adopted atmosphere parameters in Table 7.  We list the [X/H] values in 
Tables 8 - 12.  We list in Table 13 Mg and O abundances (derived from the 
9250 \AA\ region) for several stars studied in previous papers in our series.

Since we have not previously used the CTIO 1.5 m telescope for spectroscopic 
studies of SWPs, we need an independent check on the zero point of the 
derived abundances for HR\,810 and HD\,1237.  To accomplish this, we 
also obtained a spectrum of $\alpha$ Cen A with this instrument.  We derive 
the following values for T$_{\rm eff}$, $\log g$, $\xi_{\rm t}$, and [Fe/H]: 
$5774 \pm 61$ K, $4.22 \pm 0.08$, $0.90 \pm 0.10$, and $+0.35 \pm 0.05$.  
This value of [Fe/H] is 0.10 dex larger than the value derived by 
Neuforge-Verheecke \& Magain (1997).  

\subsection{Derived Parameters}

We have determined the masses and ages in the same way as in Paper V.  Using 
the {\it Hipparcos} parallaxes (ESA 1997) and the stellar evolutionary 
isochrones of Schaller et al. (1992) and Schaerer et al. (1993), along with 
our spectroscopic $T_{\rm eff}$ estimates, we have derived masses, ages, 
and theoretical $\log g$ values (Table 14).\footnote{Note, the theoretical 
$\log g$ values are derived from theoretical stellar evolutionary 
isochrones at the age which agrees with the observed T$_{\rm eff}$, 
M$_{\rm v}$, and [Fe/H] values.}  BD-10 3166 is too distant for a 
reliable parallax determination, so it is not included in the table.

Two stars, HD\,37124 and HD\,46375, give inconsistent results: they are 
located in a region of the HR diagram where no ordinary stars are expected 
(they are too luminous and/or too cool relative to even the oldest 
isochrones).  One possible solution is to invoke an unresolved companion 
of comparable luminosity.  It is highly unlikely that the companion is 
responsible for the observed radial velocity variations in each star, as 
that would require them to be viewed very nearly pole-on, which is extremely 
improbable (Geoff Marcy, private communication).  It is more likely that the 
companions are sufficiently separated such that they do not significantly 
affect the Doppler measurements on short timescales.  Therefore, we encourage 
that these two systems be searched for close stellar companions.

Several other stars, HD\,1237, HD\,130322, and HD\,192263, are of too low a 
luminosity to derive reliable ages, due to the convergence of the stellar 
evolutionary tracks at low luminosities (see Figure 1).  However, it is 
still possible to derive useful mass and $\log g$ estimates for them.  For 
those stars with theoretical $\log g$ estimates in Table 14, there is 
generally good agreement with the spectroscopic values listed in Table 7.

\section{DISCUSSION}
\subsection{Comparison with Other Studies}

Several stars in the present study have been included in other recent 
spectroscopic studies.  Santos et al. (2000b,c) analyzed a total of 13 SWPs 
using a method patterned after that of Paper V.  Two of their stars, HD\,1237 
and HD\,52265, overlap with our present sample (and one other, HD\,75289, 
from Paper V).  Their results for HD\,75289 are nearly identical to ours.  Our 
McDonald spectrum of HD\,52265 yields similar results to those of Santos et 
al. (2000b), but our Keck spectrum yields a T$_{\rm eff}$ value 100 K larger 
than theirs.  Our T$_{\rm eff}$ estimates for HD\,1237 are very similar to 
theirs, and the other parameters agree less well but are still consistent 
with our results.\footnote{Note, the youth and activity level of HD\,1237 
make it likely that this star is variable.}  Their [Fe/H] estimate for 
HD\,1237 is 0.06 dex smaller than ours.  Combining this with the results of 
our analysis of $\alpha$ Cen, we tentatively suggest that our abundance 
determinations for HR\,810 and HD\,1237 (as listed in Tables 10 and 11) be 
reduced by 0.05 dex.

When comparing our results to those of Santos et al. (2000b,c), it should be 
noted that our quoted uncertainties are smaller than theirs.  This cannot be 
due to differences in the way we calculate uncertainties, since they adopt 
the same method we employ.  Also, our EW measurements for HD\,52265 and 
HD\,HD75289 are essentially the same within 1-2 m\AA.  We suggest that a 
contributing factor is the small number of low excitation Fe I lines employed 
by them.  Clean Fe I lines with $\chi_{\rm l}$ values near 1 eV are far less 
numerous than the high excitation Fe I lines.  Adding even 2 or 3 more Fe I 
lines with small $\chi_{\rm l}$ values significantly increases the leverage 
one has in constraining T$_{\rm eff}$.

Another issue of possible concern with the Santos et al. (2000b,c) studies 
is the systematically large values of $\log g$ that they derive.  Several 
of their estimates are near 4.8.  This is 0.2 to 0.3 dex larger than is 
expected from theoretical stellar isochrones.

Abudance ratios, as [X/Fe], have also been derived by Santos et al. 
(2000b,c).  Comparing [Si/Fe], [Ca/Fe], and [Ti/Fe] values for the three 
stars in common between Santos et al. (2000b) and the present work, we find 
the results to be consistent and well within the quoted uncertainties.

Feltzing \& Gustafsson (1998) derived [Fe/H] $= +0.36$ for HD\,134987, only 
0.04 dex greater than our estimate.  Randich et al. (1999) derived [Fe/H] 
$= +0.30$ and Sadakane et al. (1999) derived [Fe/H] $= +0.31$ for HD\,217107, 
both consistent with our estimate of [Fe/H] $= +0.36$.  Fuhrmann (1998) 
derived [Fe/H] $= +0.02$ for HD\,16141, smaller than our estimate of [Fe/H] 
$= +0.15$.  Edvardsson et al. (1993) derived [Fe/H] $= +0.18$ for HD\,89744, 
smaller than our estimate of [Fe/H] $= +0.30$.  Mazeh et al. (2000) derived 
[Fe/H] $= 0.00$ for HD 209458, very close to our estimate of [Fe/H] $= 
+0.04$.  Castro et al. (1997), using a spectrum with a S/N ratio of 75, derived 
[Fe/H] $= +0.50$ for BD-10 3166, 0.17 larger than our estimate; given the 
relatively low quality of their spectrum compared to ours, we are inclined 
to consider our estimate as more reliable for this star.

Gimenez (2000) derived T$_{\rm eff}$ and [Fe/H] values for 25 SWPs from 
Str\"omgren photometry.  Nine stars are in common between the two studies, 
the results being in substantial agreement.\footnote{For HD\,168443 Gimenez 
lists T$_{\rm eff} = 6490$ K.  This is a typo; it should have been listed as 
5490 K in his paper (Alvaro Gimenez, private communication).}  In summary, 
then, our results are consistent with those of other recent studies.

\subsection{Looking for Trends}

The present total sample of SWPs with spectroscopic analyses is more than 
twice as large as that available in Paper V.  Therefore, we will make a more 
concerted effort than in our previous papers to search for trends among the 
various parameters of SWPs.  The first step is the preparation of the SWPs 
sample.

We will restrict our focus to extrasolar planets with minimum masses less 
than 11 M$_{\rm J}$.  This excludes HD\,10697 (see Zucker \& Mazeh 2000) and 
HD\,114762.  We must also exclude BD-10 3166, as it was added to Doppler 
search programs (Butler et al. 2000) as a result of our suggestion (in 
Paper IV), based on its similarity to 14\,Her and $\rho^{1}$\,Cnc.  The 
planet around HD\,89744 was also predicted prior to its announcement in 
January 2000 (see Gonzalez 2000), but it was already being monitored for 
radial velocity variations (Robert Noyes, private communication), so we will 
retain it in the sample.  The remaining stars are drawn from the previous 
papers in our series as well as the studies of Santos et al. (2000b,c), which 
are patterned after Paper V.  The total number of SWPs in the sample is 38.

We will compare the parameters of SWPs to those of field stars without known 
giant planets.  Of course, the comparison is not perfect given: 1) the 
possible presence of giant planets not yet discovered in the field star 
sample, 2) possible systematic differences between our results and those of 
the field star surveys, and 3) the possibility that some stars without known 
planets have lost them through dynamical interactions with other stars in 
its birth cluster (Laughlin \& Adams 1998).

\subsubsection{Young SWPs}

The observed metallicity distribution among nearby dwarfs is due to a 
combination of several factors: 1) the spread in age (combined with the disk 
age-metallicity relation), 2) radial mixing of stars born at different 
locations in the disk (combined with the Galactic disk radial metallicity 
gradient), and 3) intrinsic (or ``cosmic'') scatter in the initial 
metallicity.  These have the effect of blurring any additional metallicity 
trends that we may be interested in studying.  The effects of the first two 
factors can be greatly mitigated if we restrict our attention to young stars 
(i.e., age less than $\sim$ 2 Gyrs), since they have approximately the same 
age and their orbits in the Milky Way have not changed very much.

Gonzalez (2000) presented a preliminary analysis of this kind.  There he 
compares the [Fe/H] values of four young stars, HR\,810, $\tau$ Boo, 
HD\,75289, and HD\,192263, to that of a field young star sample and finds 
that all four are metal-rich relative to the mean trend in the field.  We 
repeat the comparison here with HR\,810, HD\,1237, HD\,13445, HD\,52265, 
HD\,75289, HD\,82943, HD\,89744, HD\,108147, HD\,121504, HD\,130322, 
HD\,169830, HD\,192263, and $\tau$ Boo (Figure 2); HD\,13445, HD\,82943, 
and HD\,169830 are from Santos et al. (2000b), and HD\,108147 and HD\,121504 
are from Santos et al. (2000c).  The age estimates for HD\,82943 and 
HD\,169830 quoted by Santos et al. (2000b), 5 and 4 Gyrs, respectively, are 
based on the Ca II emission measure, which is not as reliable for F stars as 
ages derived from stellar isochrones.  Ng \& Bertelli (1998) derive an age 
and mass of $2.1 \pm 0.2$ Gyrs and $1.39 \pm 0.01$ M$_{\odot}$, respectively, 
for HD\,169830; we estimate an age of $2.4 \pm 0.3$ Gyrs, based on the 
T$_{\rm eff}$ and [Fe/H] estimates of Santos et al. (2000b).  For HD\,82943 
we derive age and mass estimates of $2 \pm 1$ Gyrs and $1.16 \pm 0.02$ 
M$_{\odot}$, respectively, from Santos et al.'s (2000b) results.  Therefore, 
the age of HD\,169830 is sufficiently close to our 2 Gyr cutoff to justify 
its inclusion it in the young star subsample.  Our age and mass estimates 
for HD\,108147 and HD\,121504 are, respectively: $1 \pm 1$ Gyr, $1.23 \pm 
0.02$ M$_{\odot}$ and $2 \pm 1$ Gyr, $1.18 \pm 0.02$ M$_{\odot}$.

These results support the trend of higher mean [Fe/H] for young stars 
reported previously, except HD\,13445 and, with a lesser deviation, 
HD\,121504; both have a mean Galactocentric distance inside the Sun's orbit.  
One possible solution to this discrepancy may be that the age of HD\,13445 
has been underestimated.  If independent evidence of a greater age is found 
for HD\,13445, then it should be removed from the young star sample.

\subsubsection{Metallicity and Stellar Mass}

The detection of a correlation between metallicity and stellar mass has been 
suggested as a possible confirmation of the ``self-pollution'' scenario 
(Papers I, II).  This is due to the dependence of stellar convective envelope 
mass on stellar mass for luminosity class V stars.  Hence, the accretion of 
a given mass of high-Z material by an F dwarf will have a greater effect on 
the surface abundances than the accretion of the same amount of material by a 
G dwarf.  Laughlin (2000), using [Me/H] and mass estimates of 34 SWPs, finds 
a significantly greater correlation between [Me/H] and stellar mass among 
the SWPs compared to a field star control sample.

Santos et al. (2000b,c) also address this question.  Their analysis differs 
from Laughlin's in that they compare [Fe/H], corrected for stellar age, to 
the convective envelope mass at two ages, $10^{7}$ and $10^{8}$ years (note, 
Laughlin's comparison to a control sample eliminates the need to correct 
for stellar age).  Santos et al. (2000b,c) find a correlation that supports 
our findings, but they are not convinced it is significant.  They are 
concerned with an observational bias that reduces detection efficiency among 
F dwarfs relative to G dwarfs, due to the higher average rotation velocities 
among F dwarfs.  However, such a bias should only affect the relative number 
of detected F dwarfs with planets, not the mean [Fe/H] of the selected 
F dwarfs.

Among the young star subsample discussed in the previous section, there is 
a weak correlation between stellar mass and excess [Fe/H] (Figure 2b), which 
we define as the offset in [Fe/H] for a given star from the mean trend line 
in Figure 2a.  The nine young SWPs with mass $> 1.0$ M$_{\odot}$ are, on 
average, 0.15 dex more metal-rich than the three low mass stars (excluding 
HD\,13445).  A least-squares fit to the full sample of young SWPs results 
in a slope of $+0.63 \pm 0.26$ dex M$_{\odot}^{\rm -1}$, R of 0.60, and 
RMS scatter of 0.20 dex; leaving out HD\,13445, we find a slope of $+0.32 \pm 
0.15$ dex M$_{\odot}^{\rm -1}$, R of 0.56, and RMS scatter of 0.11 dex.  
Laughlin finds a slope of 0.548 dex M$_{\odot}^{\rm -1}$ from his dataset.

\subsubsection{Metallicity and Stellar Temperature}

The two most metal-rich SWPs, $\rho^{\rm 1}$ Cnc and 14\,Her, have similar 
temperatures, $\sim$5250 K.  Is it just a coincidence that two similar stars 
with the highest known [Fe/H] values in the solar neighborhood have planets?  
Another interesting pair is 51\,Peg and HD\,187123, which not only have 
similar atmospheric parameters but also similar planets.  On the other hand, 
HD\,52265 and HD\,75289 are virtually identical, but their planets have 
very different properties.  At the present time such comparisons are not 
very useful given the small sample size, but they will eventually help in 
isolating environmental factors not directly related to the stellar 
parameters.  For example, to account for the different parameter values of 
the planets orbiting HD\,52265 and HD\,75289, one could invoke stochastic 
planet formation mechanisms, dynamical interactions among giant planets, or 
perturbations by other stars in the birth cluster.

\subsubsection{Metallicity Distribution}

Gonzalez (2000) and Santos et al. (2000b,c) have shown that the [Fe/H] 
distribution of SWPs peaks at higher [Fe/H] than that of field dwarfs.  In 
Figure 3a we show the [Fe/H] distribution of the present sample of 38 SWPs 
with spectroscopic [Fe/H] values and compare it to the field star 
spectroscopic survey of Favata et al. (1997).\footnote{Note, we retain 62 
single G dwarfs (with T$_{\rm eff} > 5250$ K) not known to have planets from 
Favata et al.'s original sample of 91 G dwarfs.  Only one star is in common 
between our studies, HD\,210277, for which the [Fe/H] estimates differ by 
only 0.02 dex.}  The mean [Fe/H] of the SWPs sample is $+0.17 \pm 0.20$, while 
the mean of our subsample from Favata et al. is $-0.12 \pm 0.25$.

Can the difference in mean [Fe/H] between the field and SWP samples be 
accounted for entirely by the anomalously high [Fe/H] values of the more 
massive SWPs?  To attempt an answer to this question, we can correct the 
[Fe/H] values of the SWPs for the apparent correlation with stellar mass 
noted above.  We have applied the following correction: 0.15 dex is subtracted 
from [Fe/H] for SWPs with mass $> 1$ M$_{\odot}$.  We present a histogram 
with the corrected [Fe/H] values in Figure 3b.  The mean for the corrected 
sample is $0.07 \pm 0.19$.

Apart from the differences in their mean [Fe/H] values, the field star and 
SWPs samples differ in their shapes (see Figure 3a).  The uncorrected SWPs 
sample is strongly asymmetric with a peak at very high [Fe/H].  The corrected 
distribution, however, looks much more symmetric (Figure 3b).  Even after 
the correction is applied, however, there remains a small peak at extremely 
high [Fe/H].

We show the corresponding cumulative distributions in Figure 4.  A 
Kolmogorov-Smirnov test applied to the distributions in Figure 4a indicates 
a probability less than 2.8x10$^{\rm -6}$ that they are drawn from the same 
population.  The same test applied to Figure 4b indicates a probability 
less than 4.9x10$^{\rm -4}$.  Therefore, assuming there are no significant 
selection biases or systematic differences, both SWP distributions are drawn 
from significantly more metal-rich populations than the field stars.

\subsubsection{Lithium Abundances}

In Paper V we presented a simple comparison of Li abundances among SWPs 
to field stars and suggested a possible correlation in the sense that the 
SWPs have less Li, all else being equal.  Ryan (2000) presents a more 
careful comparison of Li abundances in SWPs and field stars, and concludes 
that the two groups are indistinguishable in this regard.  The present 
results do not change this conclusion.

\subsubsection{Carbon and Oxygen Abundances}

In Paper V we presented preliminary evidence for a systematic difference 
in the [C/Fe] values for SWPs relative to the Gustafsson et al. (1999) plus 
Tomkin et al. (1997) field star samples.  Most SWPs appeared to have [C/Fe] 
values less than field stars of the same [Fe/H].  The particularly low value 
of [C/Fe] for $\tau$\,Boo, a metal-rich F dwarf, lead us to select HD\,89744 
as a possible SWP that should be monitored for Doppler variations based on 
its low [C/Fe].\footnote{For additional details concerning this prediction 
and the annoucement of a planet orbiting HD\,89744, see Gonzalez (2000).}

In Paper V we did not consider possible systematic offsets among the various 
C abundance studies.  To properly compare our results to other studies, it 
is essential to determine the relative offsets.  Gustafsson et al. noted a 
systematic difference between their C abundances and those of Tomkin et al. 
(1995).  Comparing the [C/Fe] estimates of the 28 stars in common between 
these two studies, we find a signficant systematic trend with T$_{\rm eff}$; 
the two sets of [C/Fe] values are equal at T$_{\rm eff} = 5826$ K, and the 
slope is 0.00041 dex K$^{\rm -1}$ (in the sense that the Gustafsson et al. 
values are larger than those of Tomkin et al. 1995).  There are 9 stars in 
common between Tomkin et al. (1995) and Tomkin et al. (1997), with the former 
study having [C/Fe] values 0.05 dex larger on average.  We only have one star 
in common with Tomkin et al. (1997), HD\,217014, where our [C/Fe] estimate is 
smaller by 0.11 dex; we assume half this difference is due to random error, 
and, hence, adopt a systematic offset of 0.05 dex.  We have applied all these 
offsets to the various sources of [C/Fe] estimates and placed them on the 
zero-point scale of our results.  We present the results in Figure 5a.  Left 
out of the plot are Gustafsson et al. stars with T$_{\rm eff} > 6400$K, since: 
1) all the SWPs in our sample are cooler than this limit, and 2) the hottest 
stars in the Gustafsson et al. study display the largest deviations relative 
to those of Tomkin et al. (1995).  Given the large systematic offset between 
Gustafsson et al. and the other studies, we decided not to apply a correction 
for location in the Milky Way, as we did in Paper V.

We cannot determine from our present analysis alone what is the source of 
the systematic trend in the Gustafsson et al. data relative to other studies.  
However, given that they employed a single weak [C I] line and their results 
are in agreement with others for stars of solar T$_{\rm eff}$ implies that 
a weak unrecognized high excitation line is blending with it.  This is always 
the danger when basing the abundance for a given element on only one weak line.

Our new comparison of [C/Fe] values between SWPs and field stars does not 
confirm the claim we made in Paper V.  Instead, the SWPs appear to have 
slightly larger values of [C/Fe], but not significantly so.  The four SWPs 
with the largest [C/Fe] values are HD\,13445, HD\,37124, HD\,168746, 
HD\,192263.

This new result for the [C/Fe] values of SWPs relative to field stars compels 
us to revisit our successful prediction of the planet orbiting HD\,89744.  That 
prediction was based on: 1) its high [Fe/H], and 2) its low [C/Fe].  With 
the elimination of the second criterion, we are left with only one reason for 
its selection.  However, HD\,89744 is also a young F dwarf, and as shown in 
Section 4.2.1, it is much more metal-rich than the trend among young 
field stars.  Therefore, our original success for this star was accidental, 
but in light of the results presented in this work, we can look back and 
understand why we were successful.

In Figure 5b we present [O/Fe] values for field stars from Edvardsson et al. 
(1993) and Tomkin et al. (1997) and for the present sample of SWPs (using our 
average O abundances from Tables 8 - 11, 13).  As we did in Paper V for C, 
we corrected the observed [O/Fe] values for a weak trend with Galactocentric 
distance (amounting to $-0.032$ dex kpc$^{\rm -1}$).  Apart from one star 
(HD\,192263, for which we did not measure the O I triplet near 9250 \AA), 
it appears that the SWPs follow the same trend as the field stars.  The 
smaller scatter among the [O/Fe] values for field stars compared to [C/Fe] 
may be indicative of the more varied sources of C in the Milky Way.

In Figure 6 we present the [C/O] values for the same stars plotted in Figures 
4a,b.  The field stars display a positive trend with [Fe/H].  The SWPs appear 
to follow the same trend.

\subsubsection{Other Light Element Abundances}

Several other light elements have well-determined abundances in solar type 
stars: Na, Mg, Al, Si, Ca, and Ti.  Three recent spectroscopic surveys of 
nearby solar type stars have produced high quality abundance datasets for 
these elements: Edvardsson et al. (1993) and Tomkin et al. (1997), Feltzing 
\& Gustafsson (1998), and Chen et al. (2000).\footnote{These will be referred 
to as ED93, FG98, and CH00, respectively.}  They are all differential fine 
abundance studies which use the Sun as the standard for the $gf-$values.  We 
will use the results of these studies, with some modifications noted below, 
to search for possible deviations among SWPs from trends in the field star 
population.

For the ED93 sample we are: 1) retaining only the higher quality ESO results 
(and excluding their McDonald-based spectra), 2) retaining only single stars, 
3) excluding known SWPs, and 4) including the results of Tomkin et al. 
(1997).  Note that although Tomkin et al. (1997) only reanalyzed nine stars 
from Edvardsson et al., all of them are metal-rich; therefore, inclusion of 
their results greatly helps the present comparison.  For the other two surveys 
we are also retaining only single stars and excluding known SWPs.  We are also excluding K dwarfs from the FG98 sample.  Our final adopted three samples, 
ED93, FG98, and CH00, contain 62, 37, and 89 stars, respectively.  Not every 
star in these samples contains a full set of light element abundance 
determinations, so the actual number of stars available for comparison of 
abundances of a given element will be less than these totals.  In order to 
reduce systematic errors, we only employ abundances derived from neutral lines 
in forming [X/Fe] values; in the following, [Ti/Fe] is shorthand for [Ti I/Fe 
I].

Although [X/Fe] values from different studies are less likely to suffer from 
systematic differences than are [Fe/H] values (due, for instance, to 
cancellation of systematic errors in EW measurement), we must confirm that 
these various studies are consistent with each other.  We can do this by 
comparing stars in common among them.  We discuss each element in turn below:

Na -- In all the studies considered here, abundances are based on the Na I 
pair near 6160 \AA.  With ED93 we share 47\,UMa, $\rho$\,CrB, and 51\,Peg.  
Our [Na/Fe] estimates are larger by 0.1 dex for 47\,UMa, $\rho$\,CrB and 
0.1 dex smaller for 51\,Peg.\footnote{We are comparing the abundance results 
from Tomkin et al. (1997) for 51\,Peg, not Edvardsson et al., who had employed 
spectra of lesser quality for this star.}  FG98 derive a [Na/Fe] value for 
HD\,134987 0.17 dex larger than our estimate.  CH00 have in common with us 
47\,UMa and HD\,75332 for which their [Na/Fe] estimates are 0.13 and 0.00 dex 
smaller than ours, respectively.  We compare all the datasets in Figure 7 
(note, the CH00 and ED93 samples are plotted on separate diagrams for 
clarity).  Both ED93 and FG98 show an upturn in [Na/Fe] above solar [Fe/H].  
The data from CH00 do not include metal-rich stars, but are similar to ED93 
for smaller [Fe/H].  The SWPs sample deviates towards smaller [Na/Fe] by 
about 0.2 dex relative to ED93 and FG98 at high [Fe/H].  This difference 
appears to be real, but additional study of the field metal rich stars would 
be very helpful.  The star with the most deviant negative [Na/Fe] is 
HD\,192263.

Mg -- This is a more difficult element to measure, given the relatively small 
number of clean weak lines available.  Also, the neutral lines employed by 
different studies are very heterogeneous.  The large scatter evident among 
the FG98 sample in Figure 8 is perhaps evidence of the difficulty in deriving 
accurate values of [Mg/Fe] for metal-rich stars.  For the stars in common, 
our [Mg/Fe] values are consistently smaller by about 0.08 dex, on average.  
The difference between the SWPs and the metal-rich field stars in the figure 
is about twice this value.  Thus, we conclude that there is evidence for a 
real difference in [Mg/Fe], but it is tentative.

Al -- There is considerable overlap in the lines employed by different studies, 
but no two adopt exactly the same linelist.  Determinations of [Al/Fe] should 
be about as reliable as those of [Na/Fe].  Our [Al/Fe] estimates are smaller 
by about 0.04 dex than ED93, about the same as FG98, and 0.23 dex smaller 
than CH00.  As can be seen in Figure 9, the SWPs are not as cleanly separated 
from the field stars as they are for [Na/Fe] or [Mg/Fe], but there is a clump 
of SWPs with [Al/Fe] values significantly smaller than the field stars.

Si -- Abundances of Si should be considered highly reliable.  It is 
represented by several high quality neutral lines, and they display weak 
sensitivity to uncertainties in $T_{\rm eff}$ and $\log g$.  Our [Si/Fe] 
estimates are about the same as ED93 and FG98, and 0.05 dex smaller than 
CH00.  The FG98 sample displays a very large scatter in [Si/Fe] values 
compared with the other studies.  Otherwise, the SWPs sample appears as a 
continuation of the field star trends to higher [Fe/H] (Figure 10).

Ca -- The Ca abundances should also be considered reliable.  Most studies 
employ at least 3 or 4 neutral Ca lines, with one or two of them in 
common.  Our [Ca/Fe] estimates are about the same as ED93, and 0.06 dex 
smaller than CH00.  We find no evidence of a significant difference between 
the SWPs and the field stars (Figure 11).

Ti --  Like Si and Ca, the Ti abundances should be reliable.  Most studies 
employ at least three neutral lines, some many more.  The overlap is usually 
only one line, though.  The scatter among the FG98 stars is larger than the 
other samples.  Our [Ti/Fe] estimates are larger by about 0.04 dex than ED93, 
smaller by 0.04 dex than FG98, and larger by 0.04 dex than CH00.  We find no 
evidence of a significant difference between the SWPs and the field stars 
(Figure 12).

In summary, there is some evidence of a real difference in [Na/Fe], [Mg/Fe], 
and [Al/Fe] between SWPs and the general field dwarf star population.  There 
do not appear to be differnces in [Si/Fe], [Ca/Fe], and [Ti/Fe].  Among the 
F type SWPs included in the present study, the [Na/Fe], [Mg/Fe], and [Al/Fe] 
values are near $-0.05$, $0.00$, and $-0.13$, respectively.  Additional high 
quality abundance analyses of metal rich field stars are required to test 
these findings.

\subsection{Sources of Trends}

In Papers I and II, we proposed two hypotheses to account for the correlation 
between metallicity and the presence of giant planets: 1) accretion of high Z 
material after the outer convection zone of the host star has thinned to a 
certain minimum mass elevates the apparent metallicity above its primordial 
value, and/or 2) higher primordial metallicity in its birth cloud makes it 
more likely that a star will be accompanied by planets.  Laughlin presents 
evidence for the first hypothesis in the form of a weak positive correlation 
between [Me/H] and stellar mass, while Santos et al. (2000b,c) finds a 
similar, though less convincing, trend with stellar envelope mass.

Our finding of a trend between [Fe/H] and stellar mass among the young SWPs 
also supports the first hypothesis.  The mean difference in [Fe/H] between the 
high mass young SWPs and those of low mass is 0.15 to 0.20 dex.  The low mass 
young SWPs have a mean [Fe/H] similar to that of the general field young star 
population, implying that self-pollution leads to less than a $\sim 0.05$ dex 
increase for these stars.  However, given the very small sample size of young 
G dwarfs with planets, this statement is not yet conclusive.  Laws \& Gonzalez 
(2000) find a difference in [Fe/H] of $0.025 \pm 0.009$ dex between the solar 
analogs 16 Cyg A and B (16 Cyg A having the larger [Fe/H]).  This small 
difference is consistent with our lack of a detection of a metallicity anomaly 
for the low mass young solar analogs (though the number statistics are still 
small).

Subgiants allow us another way of distinguishing the two hypotheses.  After a 
solar type star leaves the main sequence and enters the subgiant branch on 
the HR diagram, the depth of its outer convection zone increases.  Sackmann 
et al. (1993) find that the Sun's outer convection zone increases by about 
0.4 M$_{\odot}$ as it traverses horizontally across the HR diagram along the 
subgiant branch (see their Figure 2 and Table 2).  Therefore, a star that has 
experienced significant pollution of its outer convection zone will undergo a 
20 fold dilution of its surface metallicity.  In this regard, HD\,38529 and 
HD\,177830 are particularly interesting.  Both stars are extremely metal rich, 
with [Fe/H] $\sim$ 0.37.  They tend to argue against the first hypothesis.  
However, the great age of HD\,177830 makes its high [Fe/H] value difficult to 
understand within either hypothesis.

As shown in the previous section, the F type SWPs appear to be deficient in 
Na and Al.  The lack of significant deviations in the C, O, Si, Ti, and Ca 
abundances among the F type SWPs is a useful clue as to the composition of 
the material possibly accreted by their host stars.  We can compare all the 
abundance anomalies to the composition of various candidate bodies.  One such 
candidate is the Earth, for which the abundances of many elements are 
relatively well known.  Using the bulk abundance estimates for the Earth 
of Kargel \& Lewis (1993), we have prepared a list of logarithmic number 
abundances relative to Fe and relative to the corresponding solar 
photospheric ratios (Table 15).  According to these estimates, C is only a 
trace element, O is moderately depleted, followed by Na, and the other 
light elements are present in roughly solar proportions.  These numbers are 
not consistent with anomalous abundance ratios we found among the F type SWPs.

Our finding of normal C/O ratios among the SWPs is inconsistent with the 
suggestion of Gaidos (2000) that a low C/O ratio is required to build giant 
planets.  Perhaps the mechanisms Gaidos discusses operate at a level 
undetectable with the present level of measurement precision.

If the anomalous F dwarfs are removed from the SWPs sample, there is still 
a significant difference relative to the field stars.  The stars HD\,12661, 
HD\,83443, HD\,134987, HD\,177830, HD\,217107, BD-10 3166, $\rho^{1}$Cnc, 
and 14\,Her have [Fe/H] values greater than 0.30 dex and T$_{\rm eff}$ 
values less than 6000 K.  Therefore, their high [Fe/H] values are more 
easily explained by the second hypothesis.  The most anomalous stars are 
still $\rho^{1}$Cnc, and 14\,Her, which do not fit easily within either 
hypothesis.

\subsection{Implications of Findings}

Assuming that a significant amount of self-pollution has indeed occurred in 
the atmospheres of the more massive SWPs, what are the implications?  There 
are several.  First, as noted in Paper II, a star with a metal-enriched 
envelope relative to its interior cannot be compared to stellar isochrones 
based on homogeneous models.  Ford et al. (1999) find that decreasing the 
interior metallicity from the observed surface value leads to a decrease in 
the derived mass but to increases in the derived age and the size of the 
convective envelope.  For 51\,Peg, Ford et al. find a decrease of 0.11 
M$_{\odot}$ in mass and an increase in age of 3.2 Gyr, if they assume an 
interior metallicity 0.2 dex less than its observed surface metallicity.  We 
encourage additional research like that of Ford et al., applied specifically 
to the F type SWPs discussed in the present work.

A possibly fruitful direction of research involves comparing the ages of the 
SWPs derived from stellar evolutionary isochrones to those obtained by other 
methods (e.g., Ca II emission measures, cluster membership, kinematics).  If 
the F type SWPs do have metal-poor interiors, then it might be possible to 
determine the systematic errors in the age estimates using such observations.

It may also be possible to learn something of the composition of the accreted 
material by comparing the deviations of the abundances of the SWPs from the 
field star population.  As was argued in the previous section, the present results are not consistent with accretion of material with the same 
composition as the Earth.

In Papers I and II we suggest that if the self-pollution mechanism is 
operating in SWPs, then we will have to adjust Galactic chemical evolution 
models accordingly, since the observed surface abundances of these stars are 
not reliable indicators of the composition of the ISM from which they 
formed.  However, the effect is likely very small since it appears that only 
F dwarfs are affected, while most Galactic chemical evolution studies employ 
abundance data from G dwarf samples.

\subsection{Brown Dwarfs}

So far only four stars have been studied spectroscopically with companions 
in the bwown dwarf mass range, 11 M$_{\rm J} <$ Mass $<$ 80 M$_{\rm J}$.  
These are HD\,10697 (present study), HD\,114762 (Paper II), HD\,162020 (Santos 
et al. 2000c), and HD\,202206 (Santos et al. 2000b) with [Fe/H] values of 
$+0.16$, $-0.60$, $+0.01$, and $+0.36$, respectively.  This is a large range, 
and, given the very small sample size, it is too early to make a meaningful 
comparison with the field stars.  Nevertheless, it is notable that two of 
these stars are quite metal-rich.  HD\,202206 is particularly interesting 
with its exceptionally high [Fe/H].

\section{CONCLUSIONS}

Employing a sample of 38 SWPs with high quality spectroscopic abundance 
analyses, we find the following anomalies:

\begin{itemize}

\item The present results confirm the high average [Fe/H] of SWPs compared 
to nearby field stars.  The average [Fe/H] of the 38 SWPs with high quality 
spectroscopic [Fe/H] estimates is $+0.17 \pm 0.20$.

\item Among the youngest SWPs, which presumably have suffered the least amount 
of migration in the Milky Way's disk, most have [Fe/H] values about 0.15 dex 
larger than young field stars at the same mean Galactocentric distance.  Young 
SWPs more massive than $\sim$1 M$_{\odot}$, in particular, have the largest 
positive ``excess [Fe/H]'' relative to the field star trend.

\item There does not appear to be significant differences in the [C/Fe] and 
[O/Fe] values between SWPs and field stars, though [C/Fe] may be 
somewhat high among the SWPs.

\item We found evidence for smaller values of [Na/Fe], [Mg/Fe], and [Al/Fe] 
among SWP's compared to field stars of the same [Fe/H].  They do not appear 
to differ in [Si/Fe], [Ca/Fe], or [Ti/Fe].

\end{itemize}

We have suggested the ``self-pollution'' scenario as an explanation for the 
anomalous trends among the F type SWPs.  However, it is not likely that this 
mechanism can account for the extremely high [Fe/H] values of 14\,Her and 
$\rho^{1}$ Cnc or of the highly evolved subgiant, HD\,177830.

Given the trends uncovered in the present work and by Laughlin, one is 
virtually guaranteed of discovering a giant planet orbiting a young F dwarf 
with a [Fe/H] value $\sim$0.25 dex greater than that of field stars at the 
same Galactocentric distance.  In this regard, we expect that a planet will 
be found orbiting HD\,75332.\footnote{Unfortunately, this star displays 
Doppler variations with an amplitude near 50 m~s$^{\rm -1}$, which appear 
to be due to chromospheric activity (Geoff Marcy, private communication).  
A high level of chromospheric activity for this star is consistent with 
our age estimate for it.}  The following four super metal rich F or G0 dwarfs 
from Feltzing \& Gustafsson (1998) also should be searched for planets: 
HD\,71479, HD\,87646, HD\,110010, and HD\,130087.  Given the trends uncovered 
in the present work, we believe the chances are high that these stars harbor 
giant planets.  It would be even more interesting if any of these stars do 
not have planets!

\acknowledgements
  
We are grateful to David Lambert for obtaining spectra of HD\,12661 and 
HD\,75332, George Wallerstein for obtaining spectra of HR\,810 and HD\,1237, 
and Geoff Marcy for sharing with us his Keck template spectra.  We thank Eric 
Gaidos for bringing to our attention the systematic offsets among the 
various C abundances studies.  Greg Laughlin and the anonymous referee also 
provided helpful comments.  This research has made use of the Simbad database, 
operated at CDS, Strasbourg, France, as well as Jean Schneider's and Geoff 
Marcy's extrasolar planets web pages.  This research has been supported by 
the Kennilworth Fund of the New York Community Trust.  Sudhi Tyagi was 
supported by the Space Grant Program at the University of Washington.

\clearpage

\section*{FIGURE CAPTIONS}

\figcaption{Locations on the HR diagram of the SWPs in the present study 
(x's); HD\,75332 is also shown ({\it open triangle}).  BD-10 3166 is not 
included in the diagram. Also shown are isochrones from Schaerer et al. (1993) 
for [Fe/H] $= +0.30$.}
 
\figcaption{Diagram {\bf a} shows [Fe/H] versus mean Galactocentric distance, 
R$_{\rm m}$, relative to the Sun's present position, R$_{\rm 0}$, for SWPs 
less than about $\sim$2 Gyrs old (x's); HD\,75332 is also 
shown ({\it open triangle}).  Also shown ({\it dots}) are field dwarfs 
fitting the same age constraint (from Gonzalez 1999).  The dotted line is a 
least-squares fit to the field stars.  Not shown in the diagram is HD\,13445, 
which has an R$_{\rm m}$ value 3.4 kpc inside the Sun's present position.  The 
large and small error bars correspond to the typical uncertainties for the 
field stars and the SWPs [Fe/H] values, respectively.  Diagram {\bf b} 
shows excess [Fe/H] plotted against stellar mass for the young SWPs from {\bf 
a}.  The datum with the large negative excess corresponds to HD\,13445.}

\figcaption{Histogram of [Fe/H] values for field stars from Favata et al. 
(1997) is shown in diagrams {\bf a} and {\bf b} ({\it shaded}).  Histogram 
of 38 SWPs is shown in bold outline in both diagrams.  Diagram {\bf a} is a 
comparison between the two raw distributions.  In diagram {\bf b} the [Fe/H] 
values of the SWPs more massive than 1.0 M$_{\odot}$ have been reduced by 
0.15 dex.}

\figcaption{Cumulative distributions of field stars ({\it dotted curve}) 
and SWP raw [Fe/H] values ({\it solid curve}) are compared in diagram 
{\bf a}.  The corresponding distribution for the corrected SWP sample is 
shown in diagram {\bf b}.}

\figcaption{Diagram {\bf a} shows [C/Fe] values for field stars from 
Gustafsson et al. ({\it filled circles}), Tomkin et al. (1997; {\it filled 
squares}), SWPs from previous studies ({\it plus signs}), SWPs from the 
present study (x's), Santos et al. ({\it open diamonds}), and HD\,75332 
({\it open triangle}). [O/Fe] values for field stars from Edvardsson et al. 
are shown in diagram {\bf b}.  Also shown are typical error bars for results 
from the present study.}

\figcaption{[C/O] values using the data from Figure 5.}

\figcaption{Shown in diagrams {\bf a} and {\bf b} are [Na/Fe] values for 
single field stars in the ED93 ({\it dots} for Edvardsson et al. and {\it 
squares} for Tomkin et al. 1997), FG98 ({\it open circles}), and CH00 ({\it 
plus signs}) samples.  Also shown are the SWPs (x's) and HD\,75332 ({\it open 
triangle}).  The error bars for the typical SWP is shown in the upper right of 
each diagram.  These comments apply also to the following five figures.}

\figcaption{[Mg/Fe] values for the samples listed in caption to Figure 7.}

\figcaption{[Al/Fe] values for the samples listed in caption to Figure 7.}

\figcaption{[Si/Fe] values for the samples listed in caption to Figure 7.}

\figcaption{[Ca/Fe] values for the samples listed in caption to Figure 7.}

\figcaption{[Ti/Fe] values for the samples listed in caption to Figure 7.}

\clearpage



\clearpage

\begin{deluxetable}{lccccc}
\tablecaption{Fe I and II lines Incorporated Since Paper V}
\scriptsize
\tablewidth{0pt}
\tablehead{
\colhead{Species} & \colhead{$\lambda_{\rm o}$} & 
\colhead{$\chi_{\rm 1}$} & \colhead{$\log gf$} & \colhead{EW$_{\odot}$}\\
\colhead{} & \colhead{(\AA)} & \colhead{(eV)} & \colhead{} & 
\colhead{(m\AA)}
}
\startdata
Fe I & 6024.07 & 4.55 & -0.12 & 118.0\nl
Fe I & 6093.65 & 4.61 & -1.34 & 31.0\nl
Fe I & 6096.67 & 3.98 & -1.81 & 37.2\nl
Fe I & 6098.25 & 4.56 & -1.74 & 17.1\nl
Fe I & 6213.44 & 2.22 & -2.66 & 83.5\nl
Fe I & 6820.37 & 4.64 & -1.17 & 42.0\nl
Fe I & 7583.80 & 3.02 & -1.90 & 85.7\nl
Fe I & 7586.03 & 4.31 & -0.18 & 135.3\nl
Fe II & 5991.38 & 3.15 & -3.48 & 30.3\nl
Fe II & 6442.95 & 5.55 & -2.38 & 5.2\nl
Fe II & 6446.40 & 6.22 & -1.92 & 4.0\nl
\enddata
\end{deluxetable}

\clearpage

\begin{deluxetable}{lcccc}
\tablecaption{[X/H] Values for HD\,1237}
\scriptsize
\tablewidth{0pt}
\tablehead{
\colhead{Element} & \colhead{n1} & 
\colhead{n2} & \colhead{n3} & \colhead{avg}
}
\startdata
Li & $+1.05 \pm 0.05$ & $+0.94 \pm 0.05$ & $+0.98 \pm 0.05$ & $+0.99 
\pm 0.03$ \nl
O(trip) & $+0.09 \pm 0.10$ & $+0.18 \pm 0.06$ & $+0.06 \pm 0.07$ & $+0.12 
\pm 0.04$ \nl
O(corr) & $+0.11 \pm 0.10$ & $+0.20 \pm 0.06$ & $+0.08 \pm 0.07$ & $+0.14 
\pm 0.04$ \nl
Na & $+0.14 \pm 0.04$ & $+0.05 \pm 0.03$ & $+0.09 \pm 0.04$ & $+0.08 
\pm 0.02$ \nl
Mg & $+0.07 \pm 0.09$ & $+0.11 \pm 0.07$ & $+0.09 \pm 0.09$ & $+0.09 
\pm 0.05$ \nl
Al(78) & $+0.09 \pm 0.04$ & $+0.00 \pm 0.04$ & $+0.13 \pm 0.05$ & $+0.07 
\pm 0.02$ \nl
Al(synth) & $+0.12 \pm 0.05$ & $+0.01 \pm 0.05$ & $+0.06 \pm 0.05$ & $+0.06 
\pm 0.03$ \nl
Al(avg) & $+0.10 \pm 0.03$ & $+0.00 \pm 0.03$ & $+0.15 \pm 0.04$ & $+0.07 
\pm 0.02$ \nl
Si & $+0.18 \pm 0.02$ & $+0.14 \pm 0.02$ & $+0.15 \pm 0.03$ & $+0.16 
\pm 0.01$ \nl
S & $+0.28 \pm 0.09$ & $+0.27 \pm 0.07$ & $+0.17 \pm 0.08$ & $+0.24 \pm 
0.05$ \nl
Ca & $+0.20 \pm 0.05$ & $+0.17 \pm 0.04$ & $+0.22 \pm 0.07$ & $+0.19 
\pm 0.03$ \nl
Sc & $+0.25 \pm 0.06$ & $+0.19 \pm 0.04$ & $+0.02 \pm 0.06$ & $+0.16 
\pm 0.03$ \nl
Ti I & $+0.24 \pm 0.08$ & $+0.11 \pm 0.06$ & $+0.11 \pm 0.07$ & $+0.14 
\pm 0.04$ \nl
Ti II & \nodata & $+0.16 \pm 0.05$ & $+0.01 \pm 0.05$ & $+0.09 
\pm 0.04$ \nl
Ti(avg) & $+0.24 \pm 0.08$ & $+0.14 \pm 0.04$ & $+0.04 \pm 0.04$ & $+0.11 
\pm 0.03$ \nl
Cr & $+0.16 \pm 0.09$ & $+0.16 \pm 0.07$ & $+0.14 \pm 0.09$ & $+0.15 
\pm 0.05$ \nl
Fe & $+0.19 \pm 0.05$ & $+0.14 \pm 0.04$ & $+0.16 \pm 0.05$ & $+0.16 
\pm 0.03$ \nl
Ni & $+0.22 \pm 0.10$ & $+0.11 \pm 0.09$ & $-0.03 \pm 0.10$ & $+0.10 
\pm 0.06$ \nl
\enddata
\end{deluxetable}

\clearpage

\begin{deluxetable}{lccccc}
\tablecaption{[O/H] and [Mg/H] values for SWPs Analyzed in Previous Papers 
in Our Series}
\scriptsize
\tablewidth{0pt}
\tablehead{
\colhead{Element} & \colhead{16 Cyg A} & \colhead{16 Cyg B} & 
\colhead{$\upsilon$ And} & \colhead{$\tau$ Boo} & 
\colhead{HD 75289}\\
}
\startdata
Mg & $+0.16 \pm 0.06$ & $+0.09 \pm 0.06$ & 
$+0.25 \pm 0.05$ & $+0.44 \pm 0.06$ & \nodata\nl
O(synth) & $-0.01 \pm 0.07$ & $+0.04 \pm 0.07$ & 
$+0.23 \pm 0.06$ & $+0.10 \pm 0.07$ & \nodata\nl
EW(7771) & 81.4 & 75.4 & 120.0 & 153.0 & \nodata\nl
EW(7774) & 70.1 & 63.3 & 105.6 & 135.3 & 95.3\nl
EW(7775) & 56.9 & 50.7 & 87.4 & 113.3 & 79.2\nl
O(trip) & $+0.10 \pm 0.06$ & $+0.06 \pm 0.06$ & 
$+0.22 \pm 0.05$ & $+0.42 \pm 0.06$ & $+0.14 \pm 0.05$\nl
O(corr) & $+0.10 \pm 0.06$ & $+0.05 \pm 0.06$ & 
$+0.12 \pm 0.05$ & $+0.25 \pm 0.06$ & $+0.06 \pm 0.05$\nl
O(avg) & $+0.05 \pm 0.05$ & $+0.05 \pm 0.05$ & 
$+0.17 \pm 0.04$ & $+0.19 \pm 0.05$ & $+0.06 \pm 0.05$\nl
C/O & $+0.09 \pm 0.05$ & $+0.10 \pm 0.05$ & 
$-0.11 \pm 0.05$ & $-0.13 \pm 0.05$ & $+0.10 \pm 0.04$\nl
\enddata
\end{deluxetable}

\clearpage

\begin{deluxetable}{lc}
\tablecaption{Composition of Earth Relative to Solar 
Photosphere\tablenotemark{a}}
\scriptsize
\tablewidth{0pt}
\tablehead{
\colhead{Element Ratio} & \colhead{[X/Fe]$_{\oplus}$} 
\\
}
\startdata
C/Fe & $-4.21$\nl
O/Fe & $-0.79$\nl
Na/Fe & $-0.56$\nl
Mg/Fe & $-0.05$\nl
Al/Fe & $+0.00$\nl
Si/Fe & $-0.09$\nl
Ca/Fe & $+0.00$\nl
Ti/Fe & $-0.05$\nl
\enddata
\tablenotetext{a}{For the purposes of calculating the numbers in this 
table, we adopted the solar phospheric abundances listed by Grevesse \& 
Sauval (1998).}
\end{deluxetable}

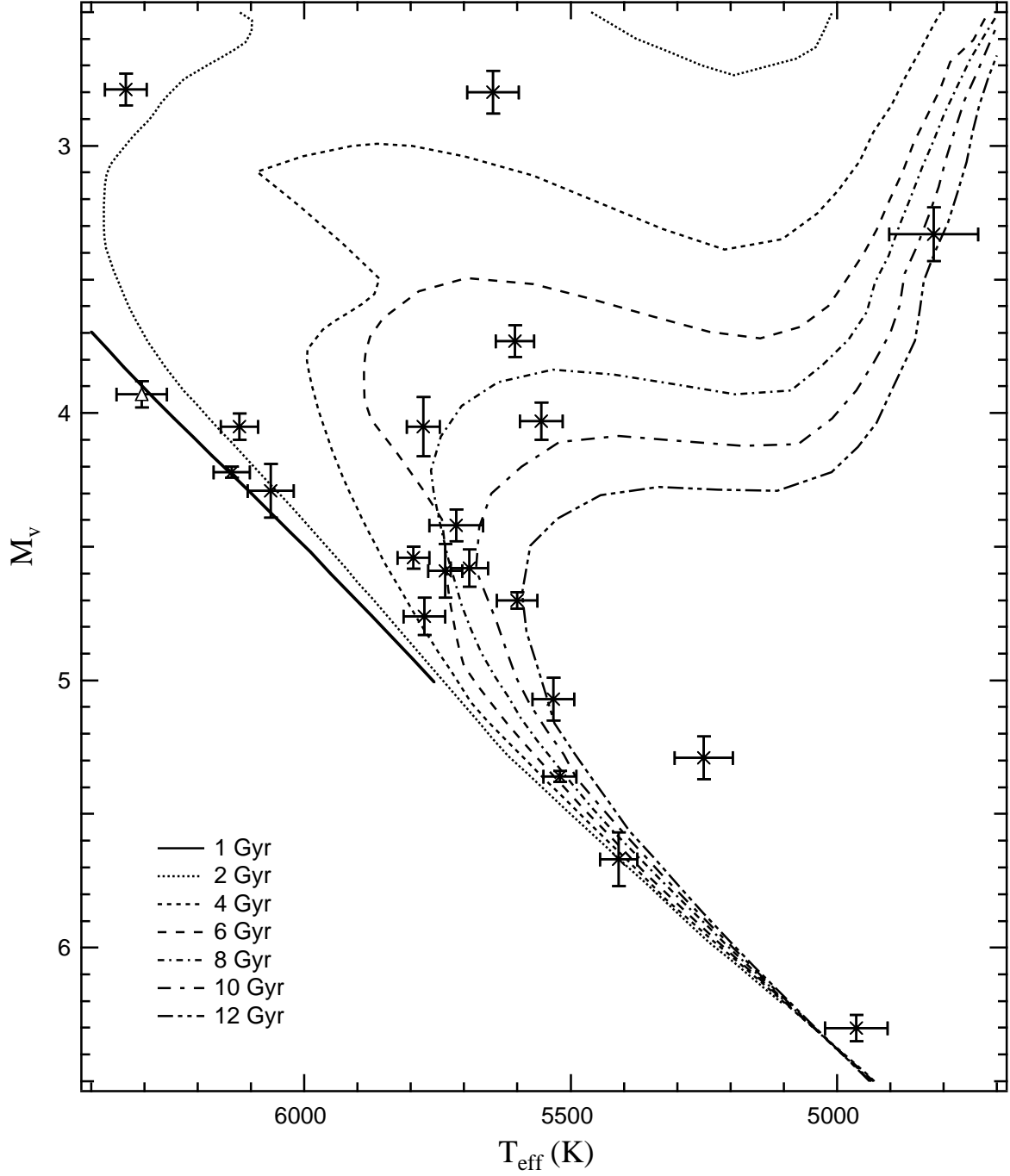
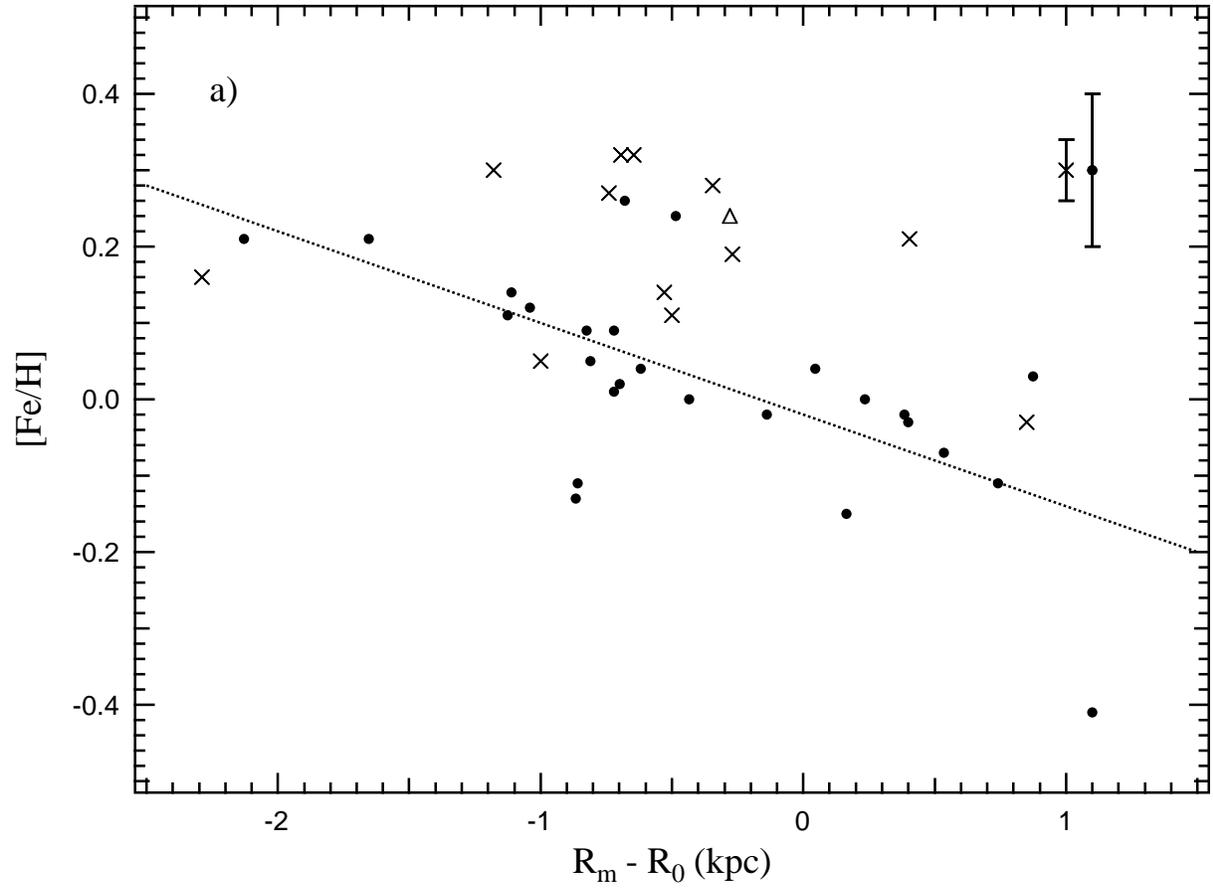
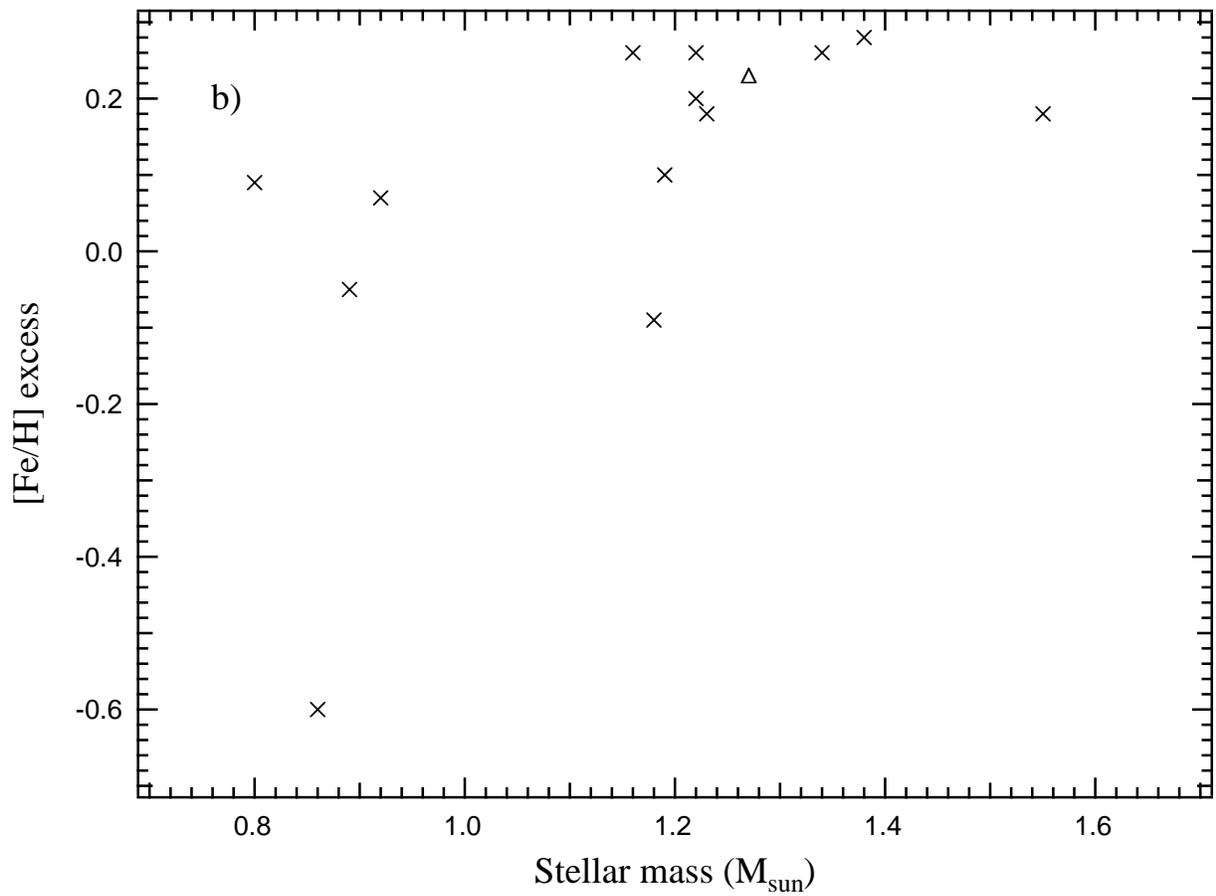
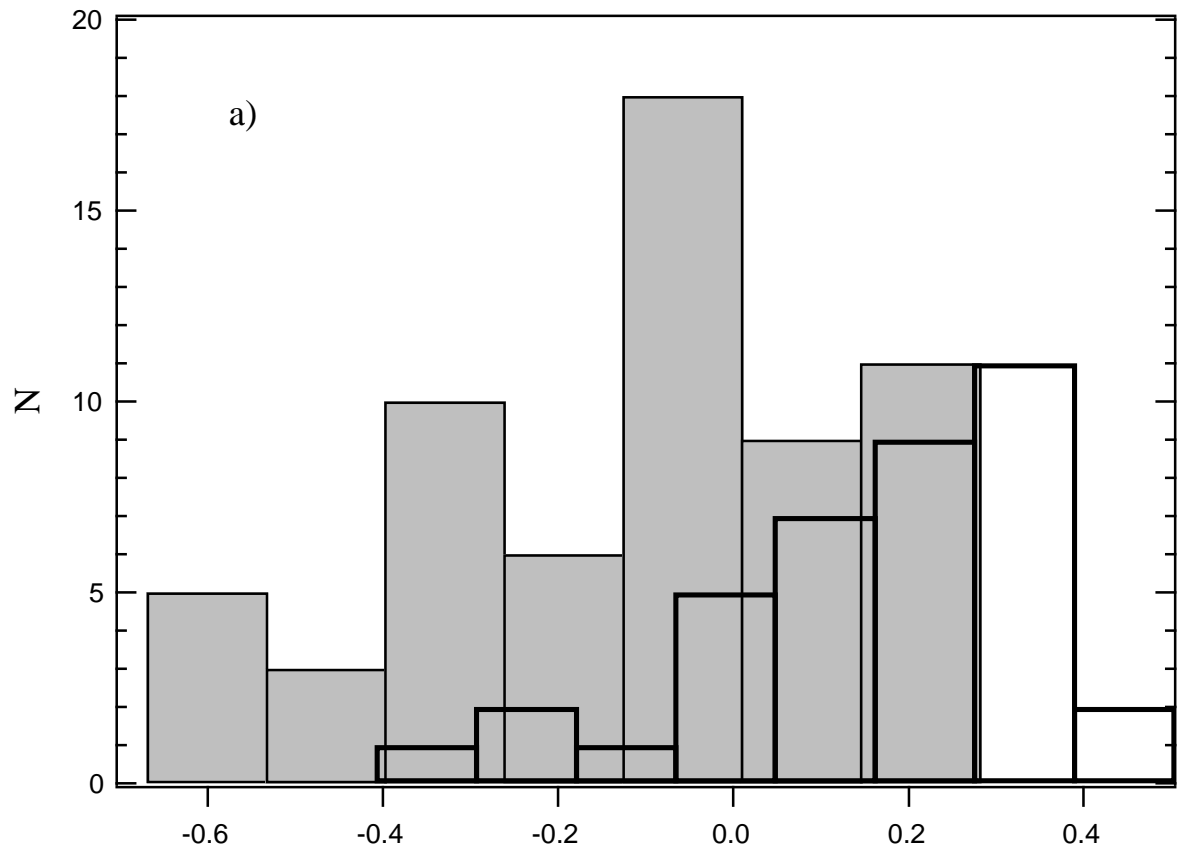
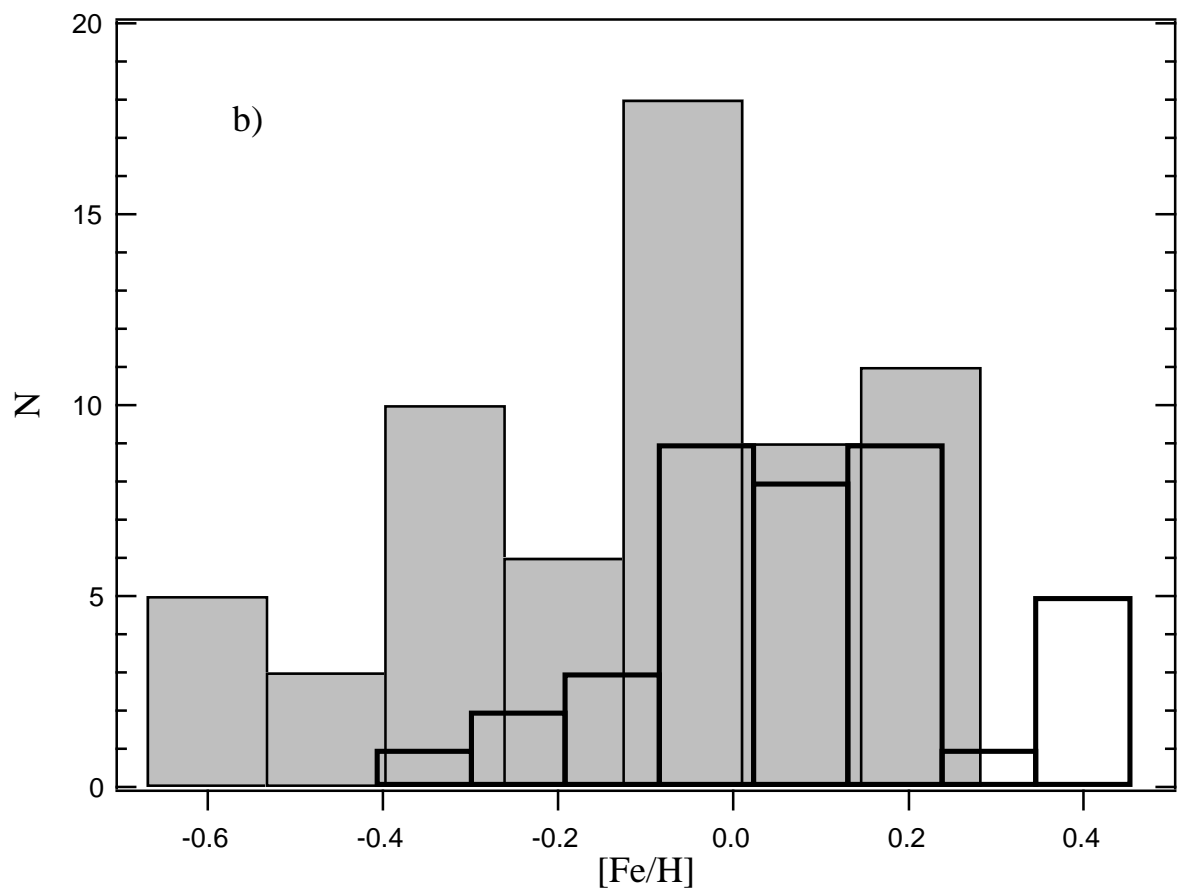
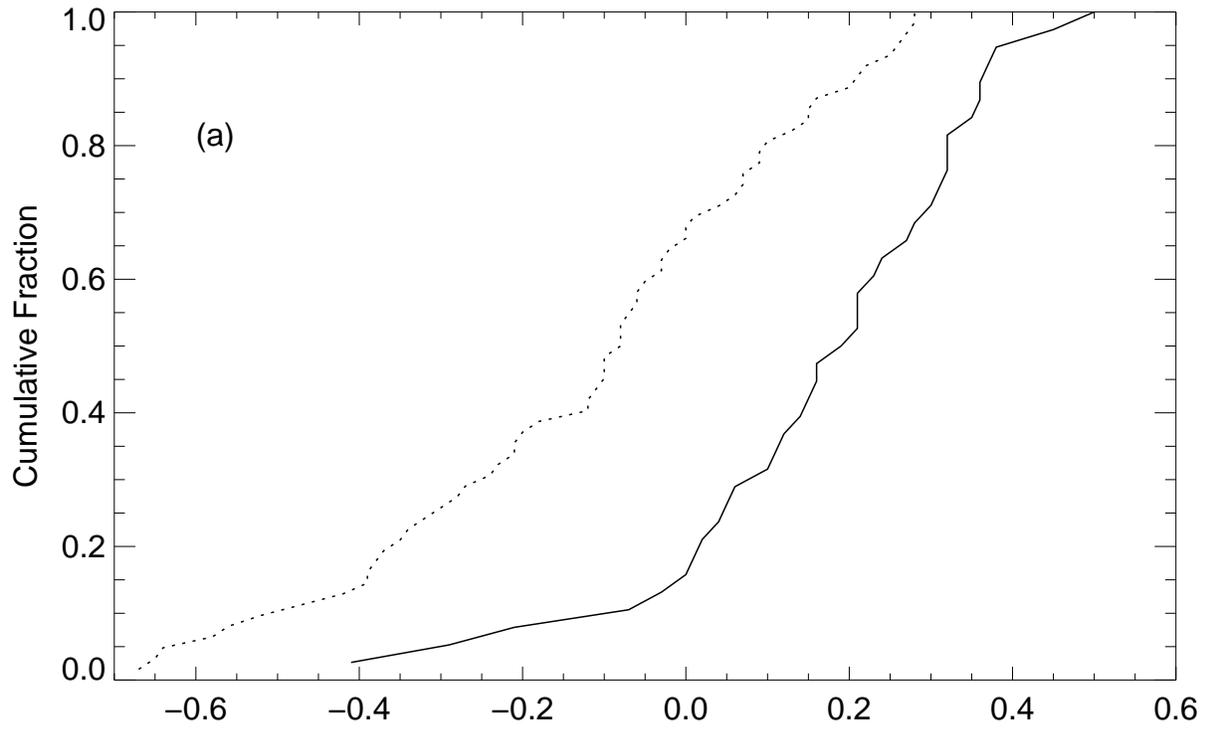
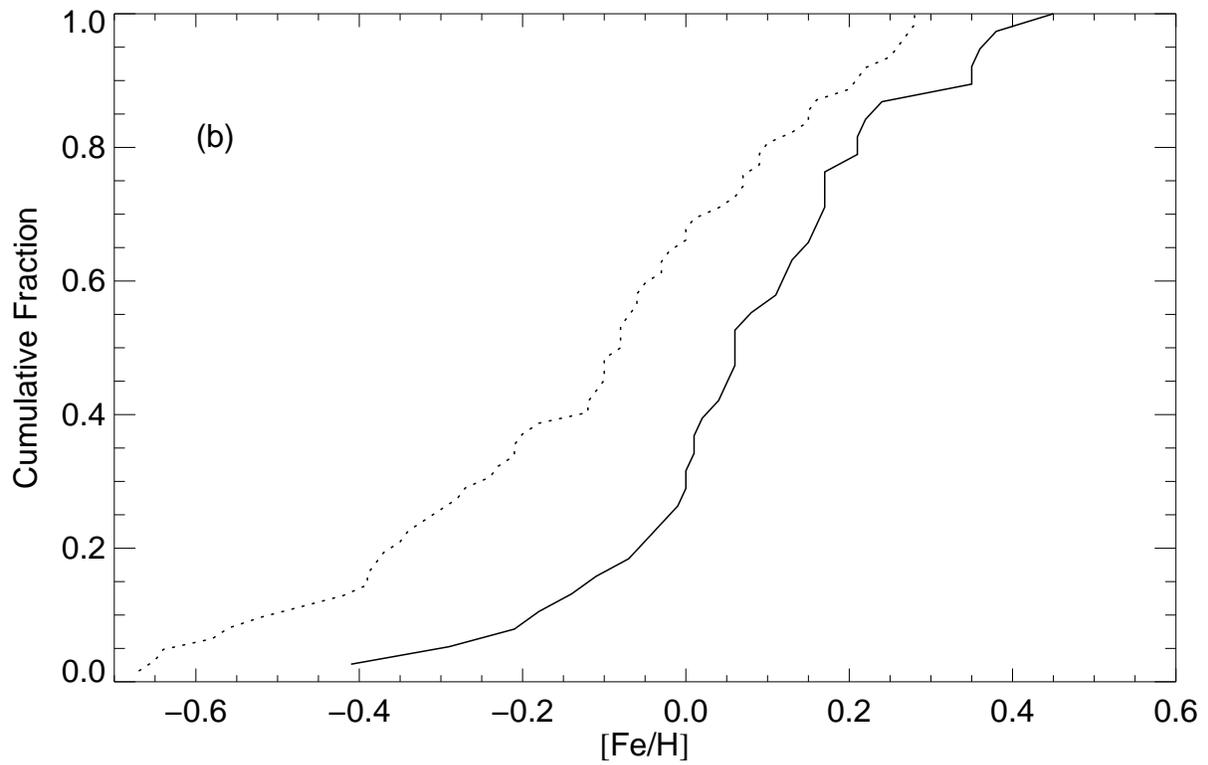
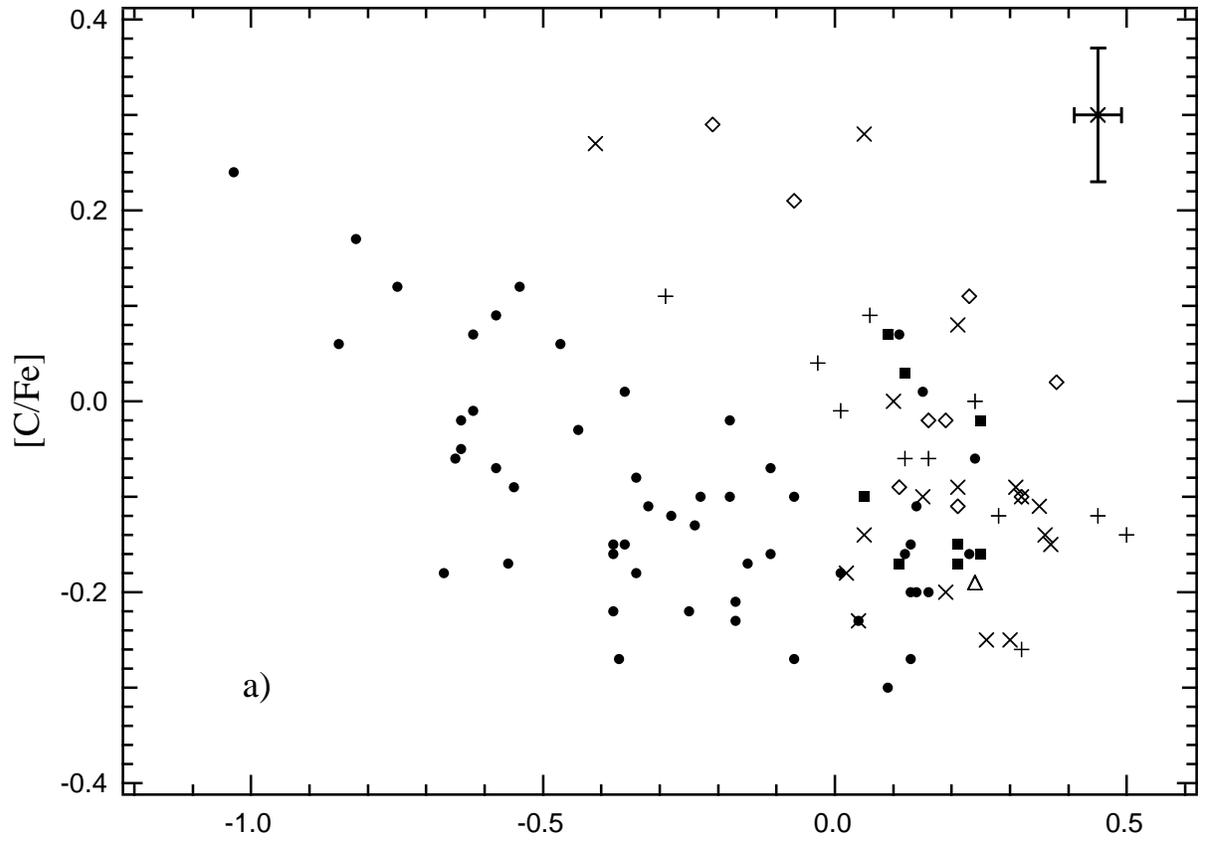
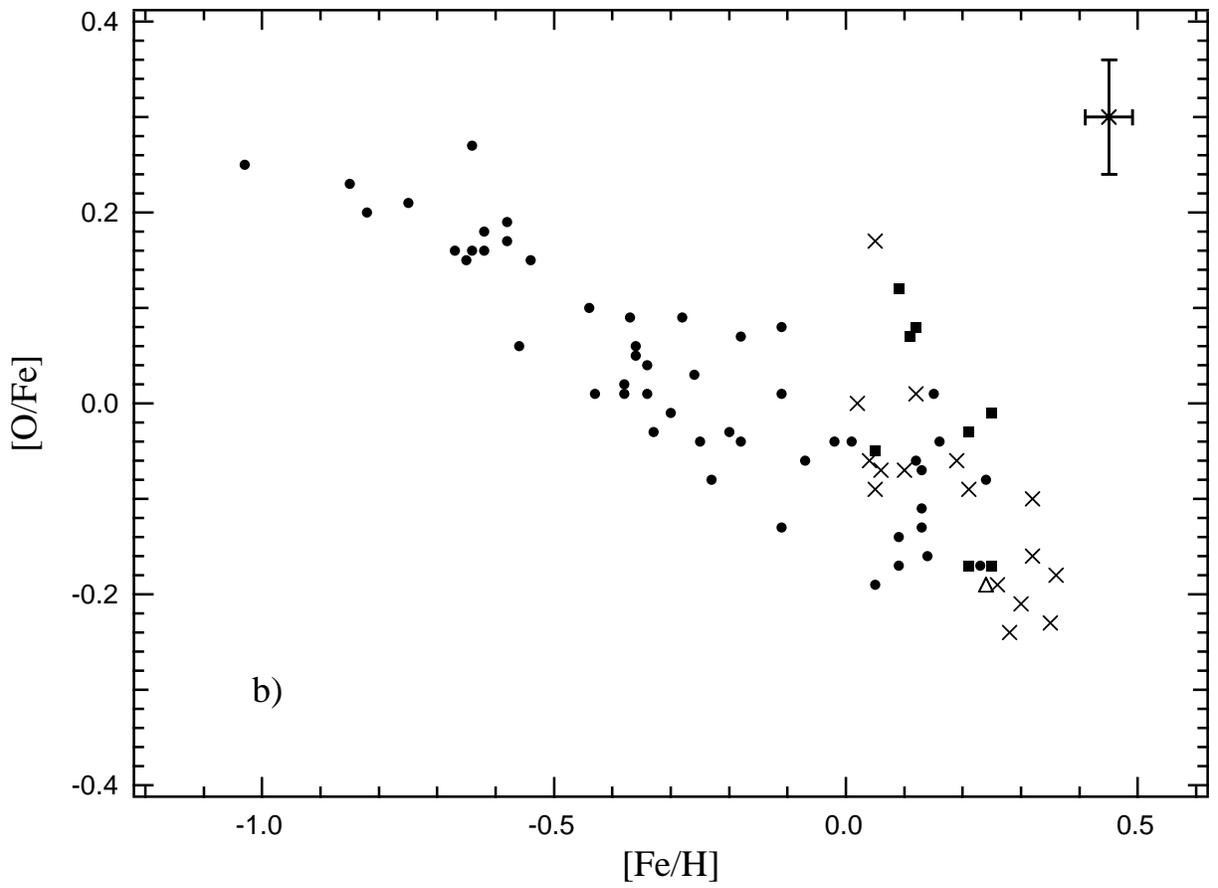
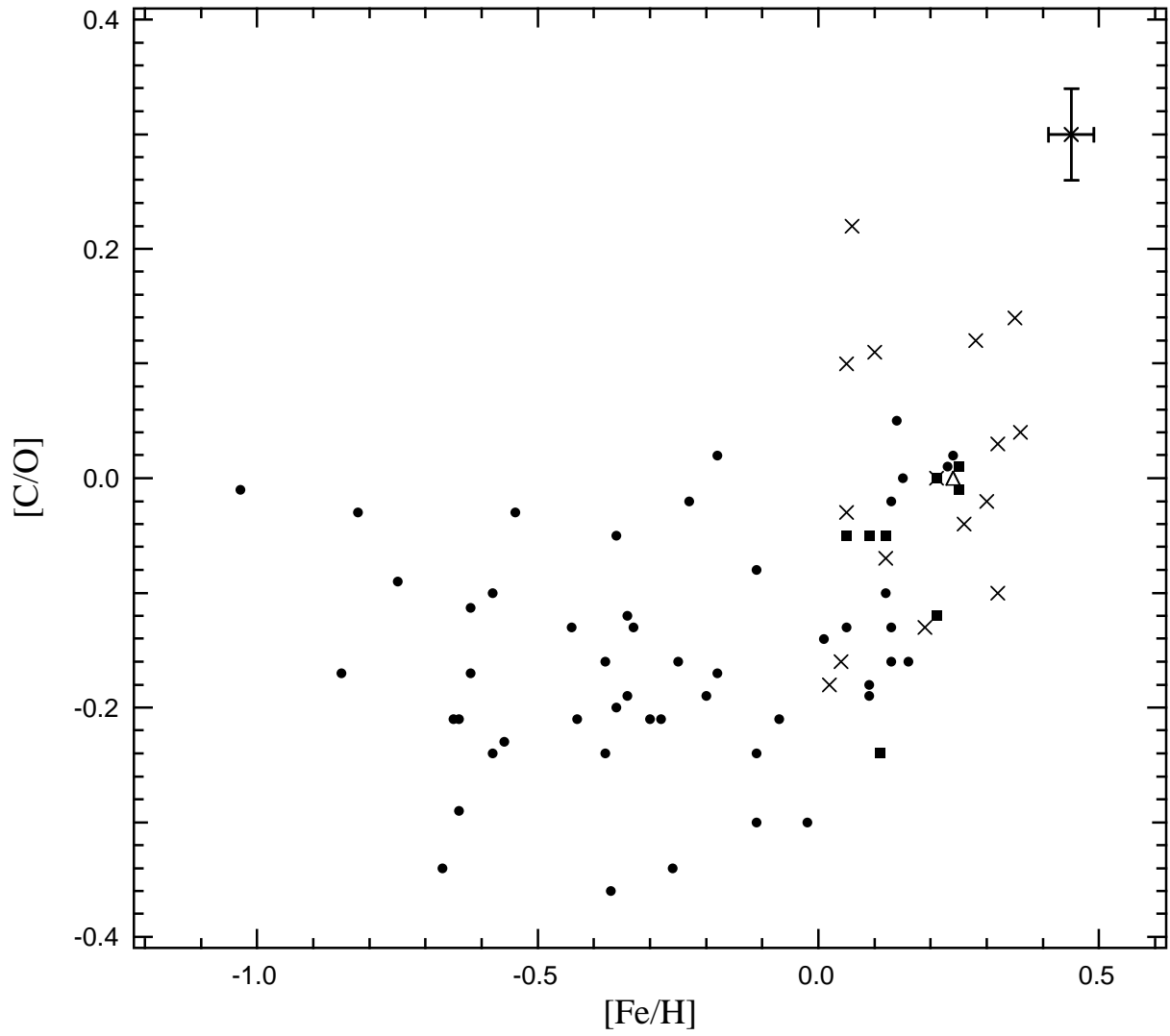
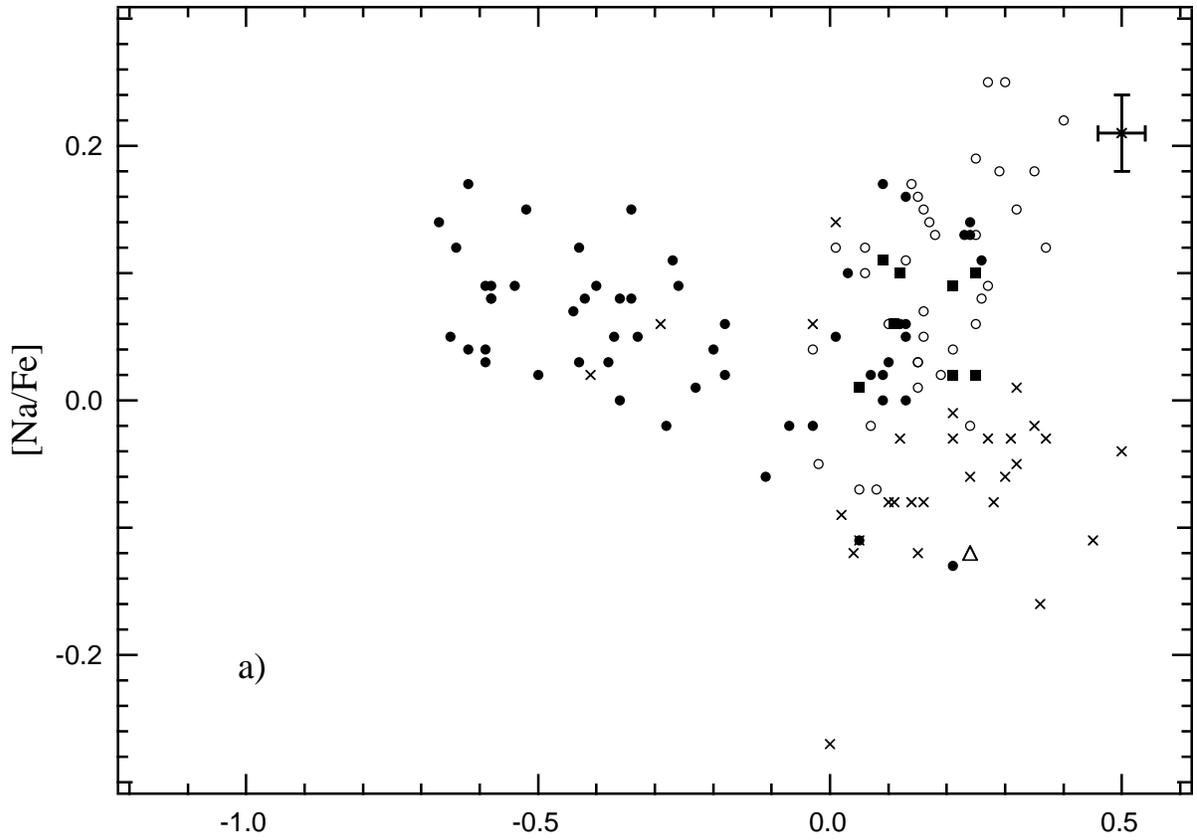
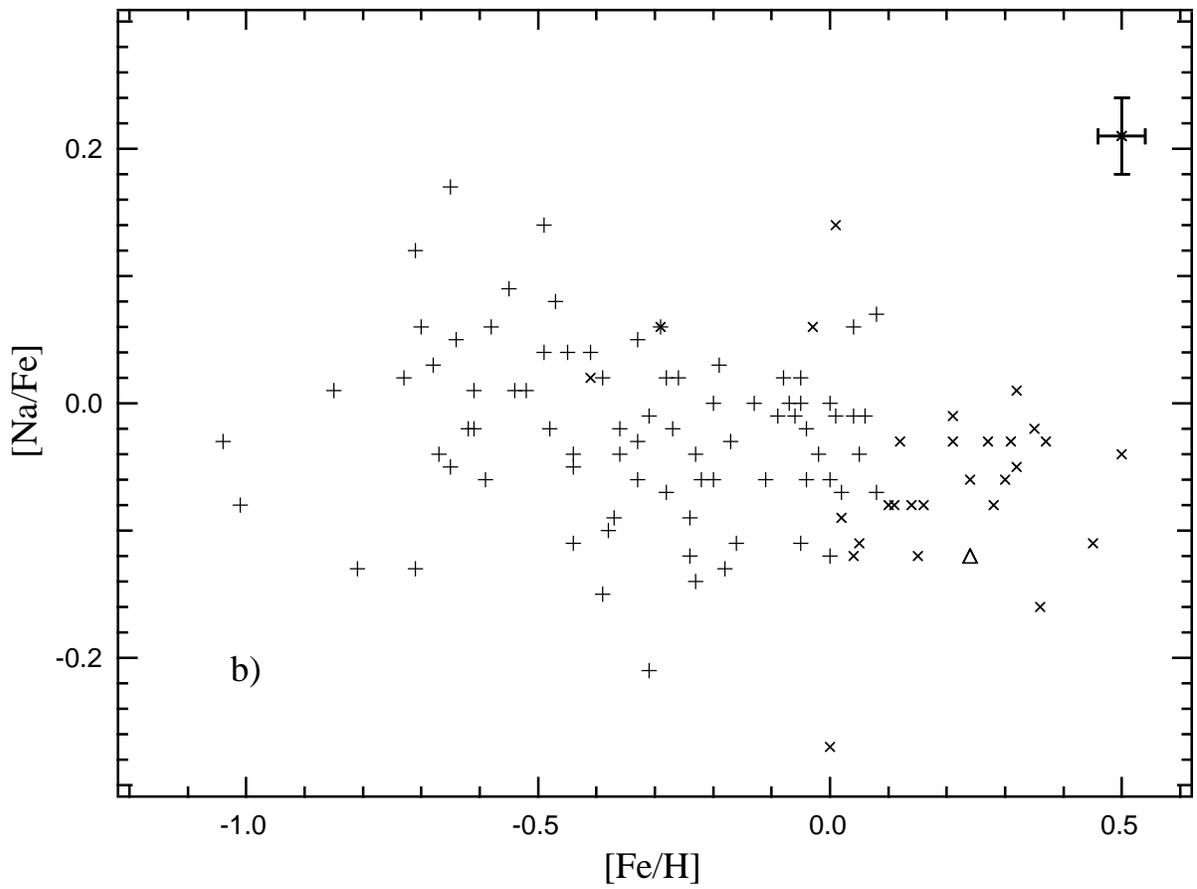
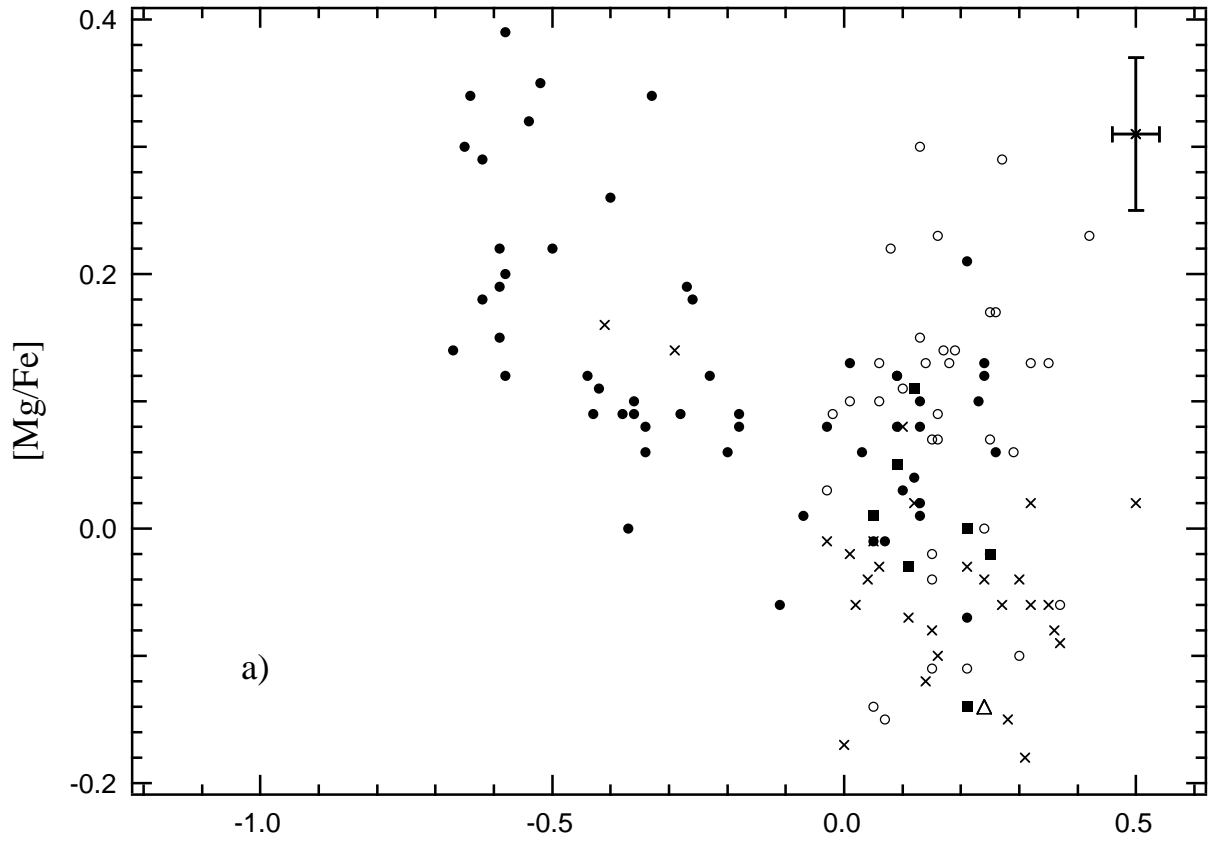
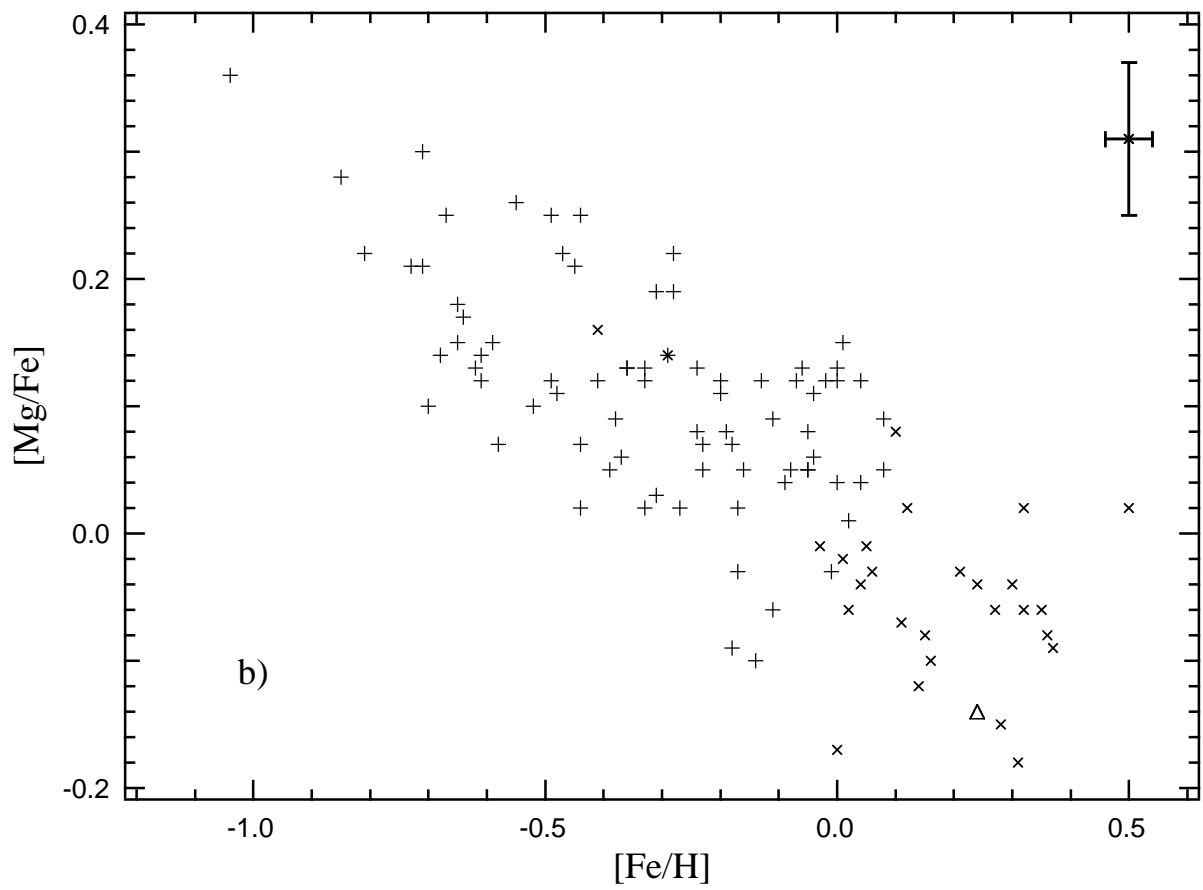
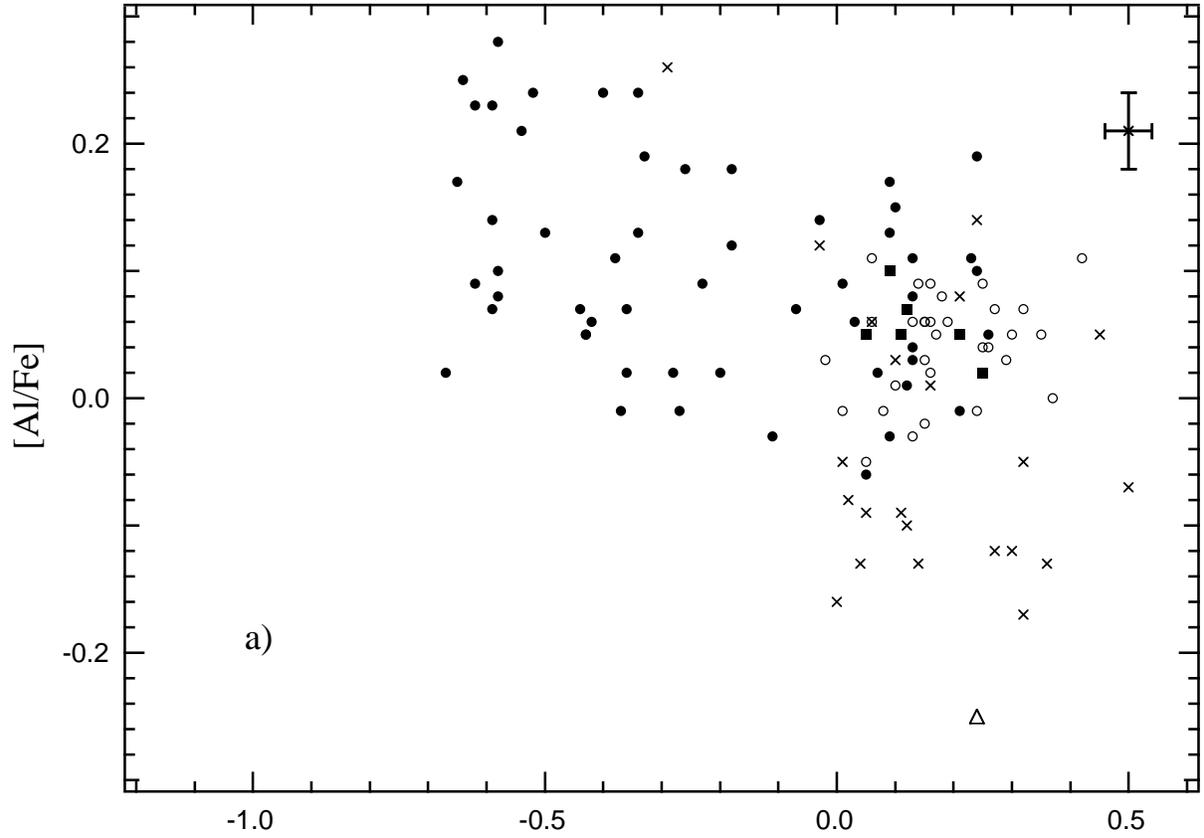
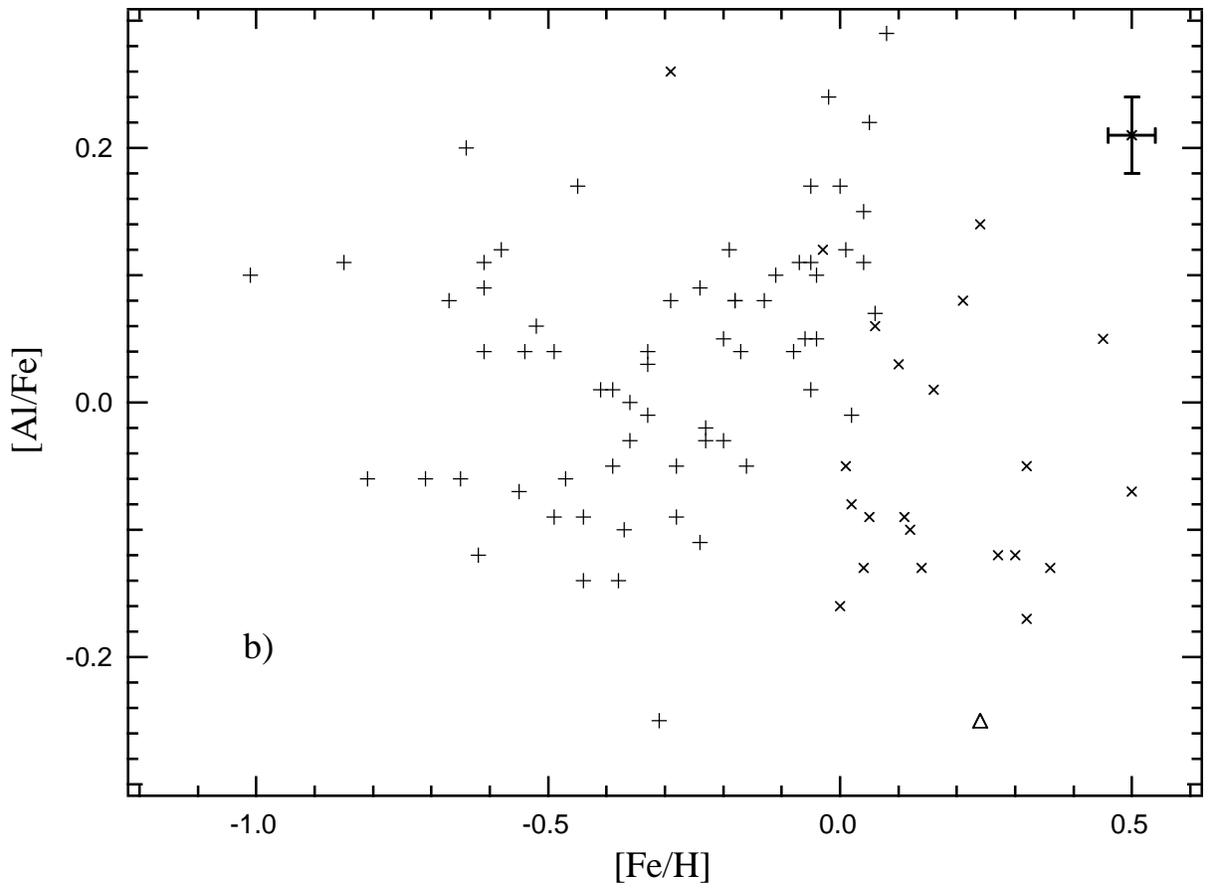
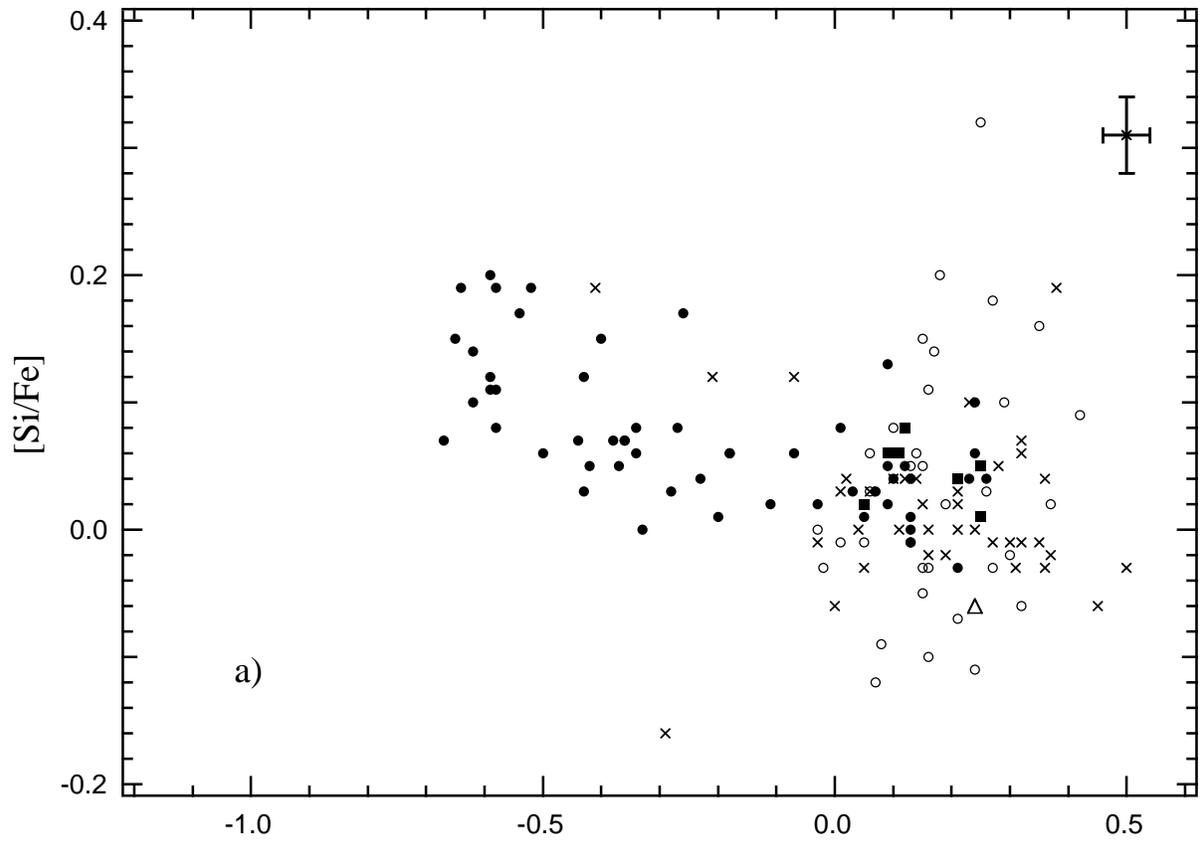
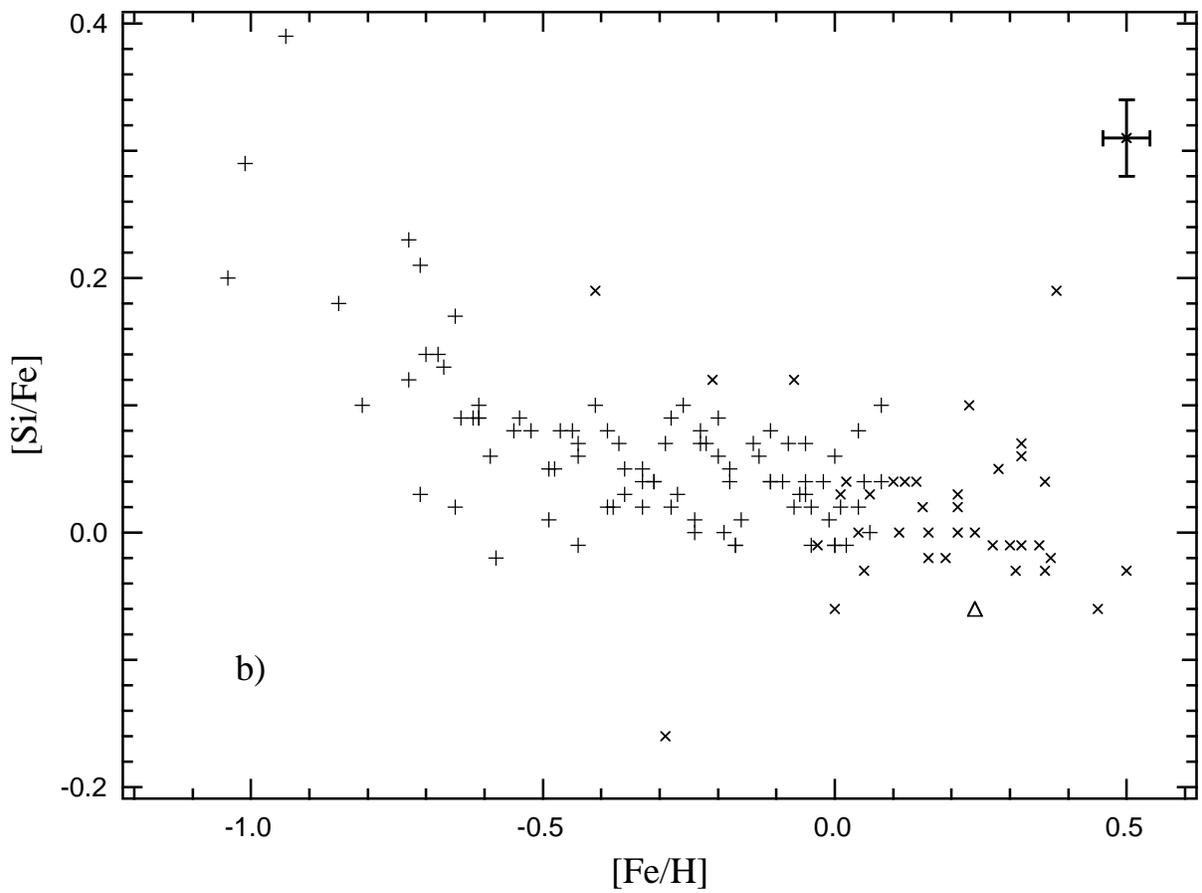
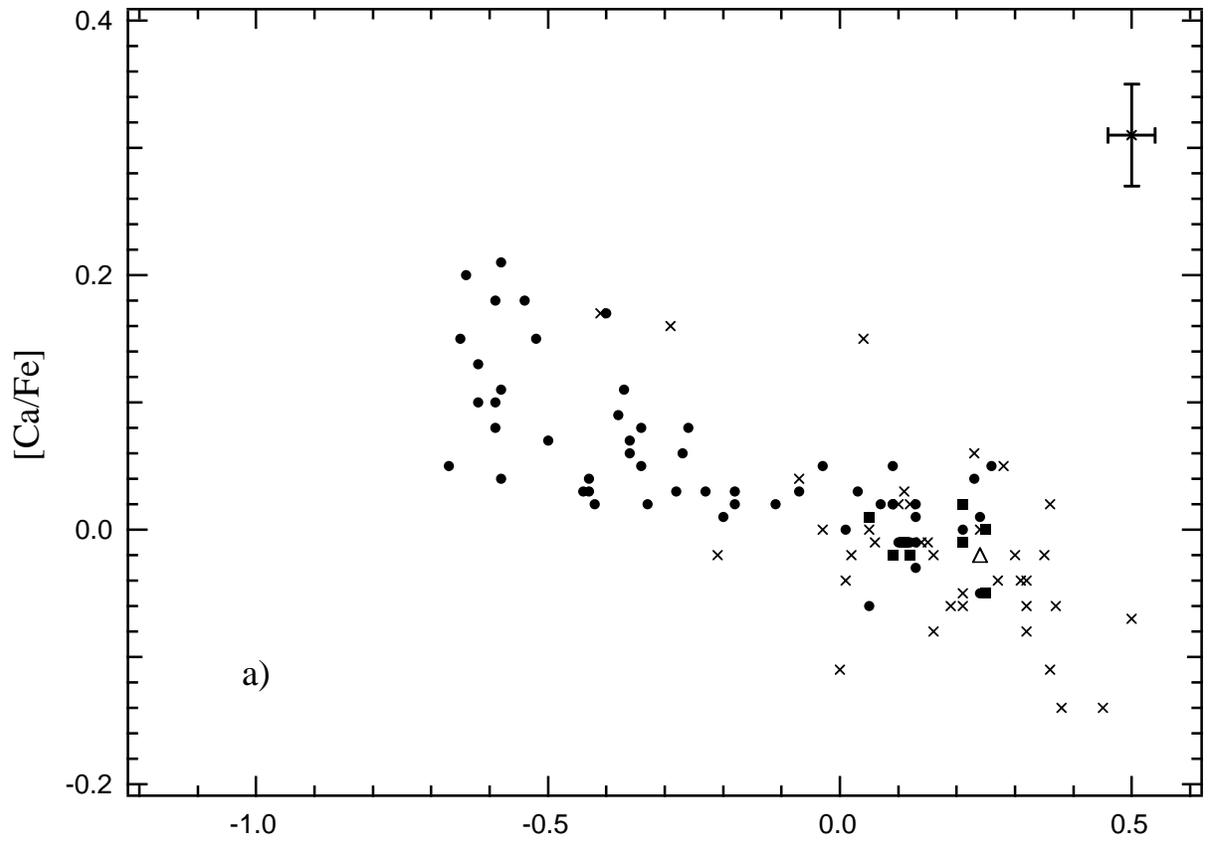
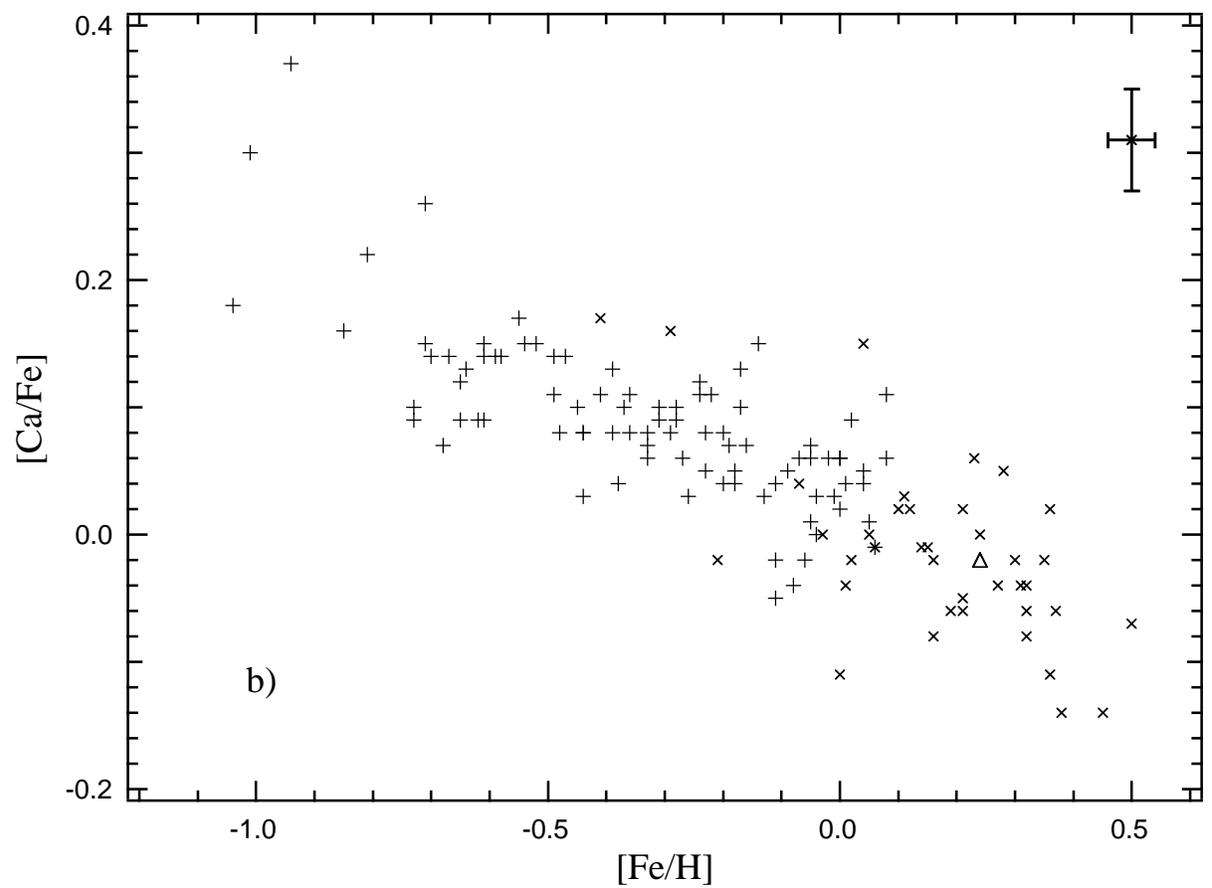
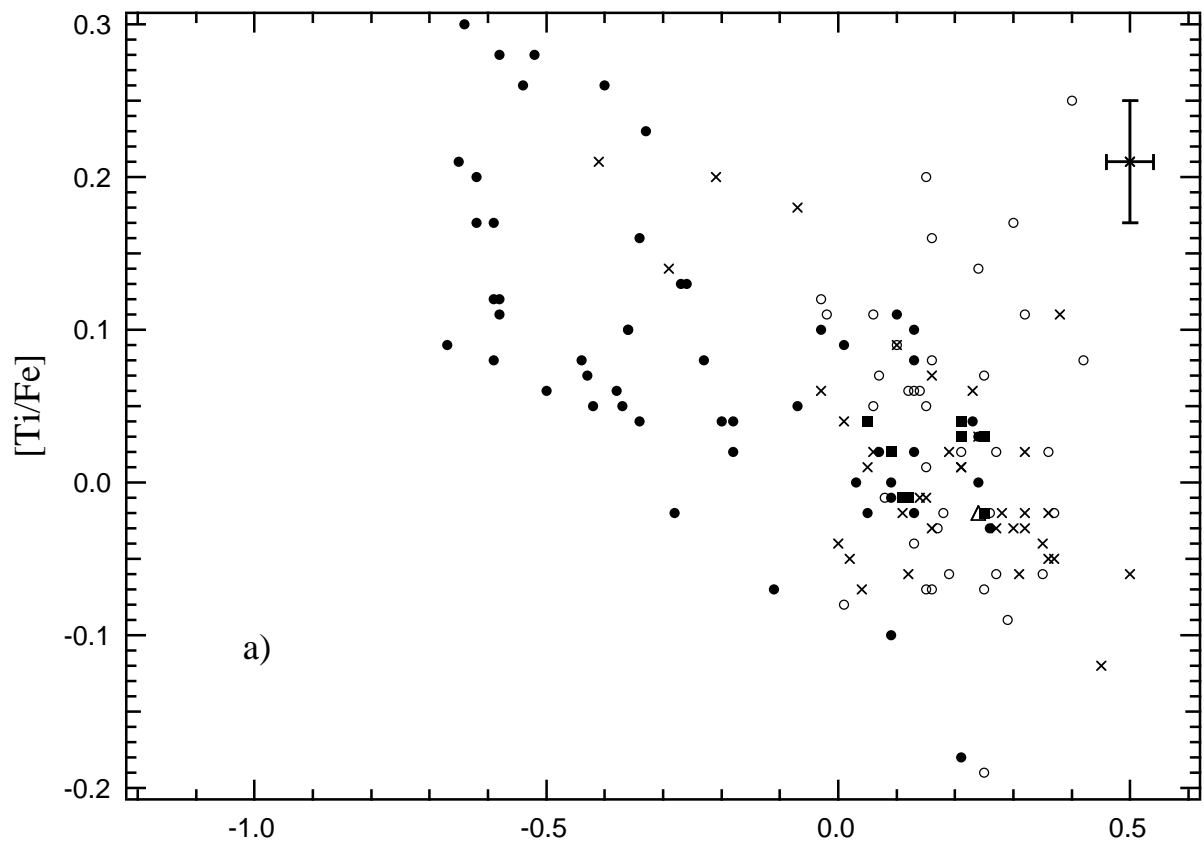
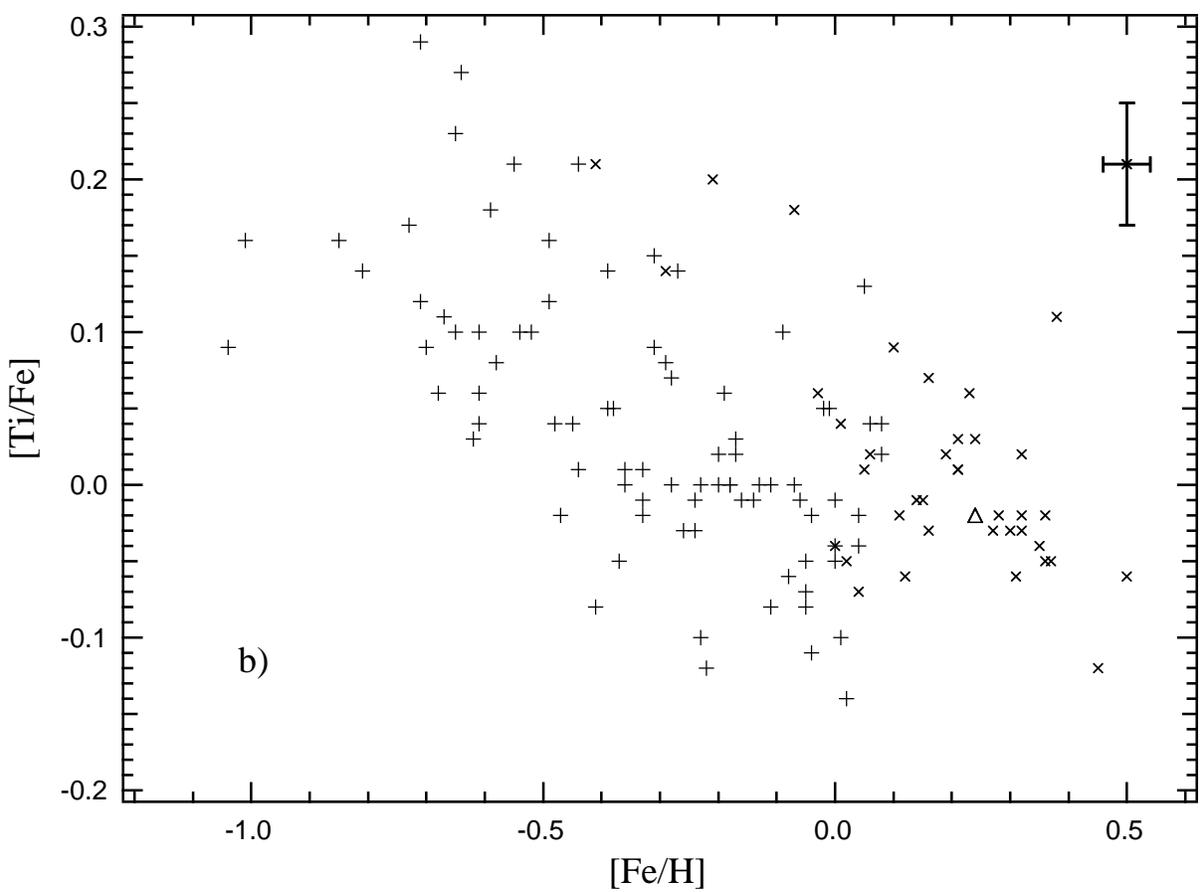
\end{document}